%% file: mainOJCYS.tex
\documentclass{IEEEojcsys}
\input{_aux/preamble}
\jvol{00}
\jnum{XX}
\paper{1234567}
\pubyear{2026}
\receiveddate{August 2026}
\begin{document}

\sptitle{Article Category}

\title{Uncertainty Estimators for Robust Backup Control Barrier Functions} 

\author{DAVID E. J. VAN WIJK\affilmark{1}}
\author{ERSIN DA\c{S}\affilmark{2}}
\author{ANIL ALAN\affilmark{3}}
\author{SAMUEL COOGAN\affilmark{4}}
\author{TAMAS G. MOLNAR\affilmark{5}}
\author{JOEL W. BURDICK\affilmark{1}}
\author{MANORANJAN MAJJI\affilmark{6}}
\author{KERIANNE L. HOBBS\affilmark{7}}

\affil{California Institute of Technology, Pasadena, CA 91125, USA} 
\affil{Illinois Institute of Technology, Chicago, IL 60616, USA} 
\affil{Delft University of Technology, Delft, The Netherlands} 
\affil{Georgia Institute of Technology, Atlanta GA 30332, USA} 
\affil{Cleveland State University, Cleveland OH 44115, USA} 
\affil{Texas A\&M University, College Station TX 77845, USA} 
\affil{The Aerospace Corporation, Chantilly, VA 20151, USA} 

\corresp{CORRESPONDING AUTHOR: D. E. J. van Wijk (e-mail: \href{mailto:vanwijk@caltech.edu}{vanwijk@caltech.edu})}
\authornote{This work was supported by in part by DARPA under the LINC program and by the Technology Innovation Institute (TII).}

\markboth{UNCERTAINTY ESTIMATORS FOR
ROBUST BACKUP CONTROL BARRIER FUNCTIONS}{D. E. J. VAN WIJK {\itshape ET AL}.}

\input{sections/01_Abstract}


\maketitle

\input{sections/02_Introduction}
\input{sections/03_Preliminaries}
\input{sections/04_MainResult}
\input{sections/05_NumericalExamples}
\input{sections/06_Conclusion}
\input{sections/07_Appendix}

\bibliographystyle{ieeetr}
\bibliography{bib/references}

\begin{IEEEbiography}
[{\includegraphics[width=1in,height=1.25in,clip,keepaspectratio]{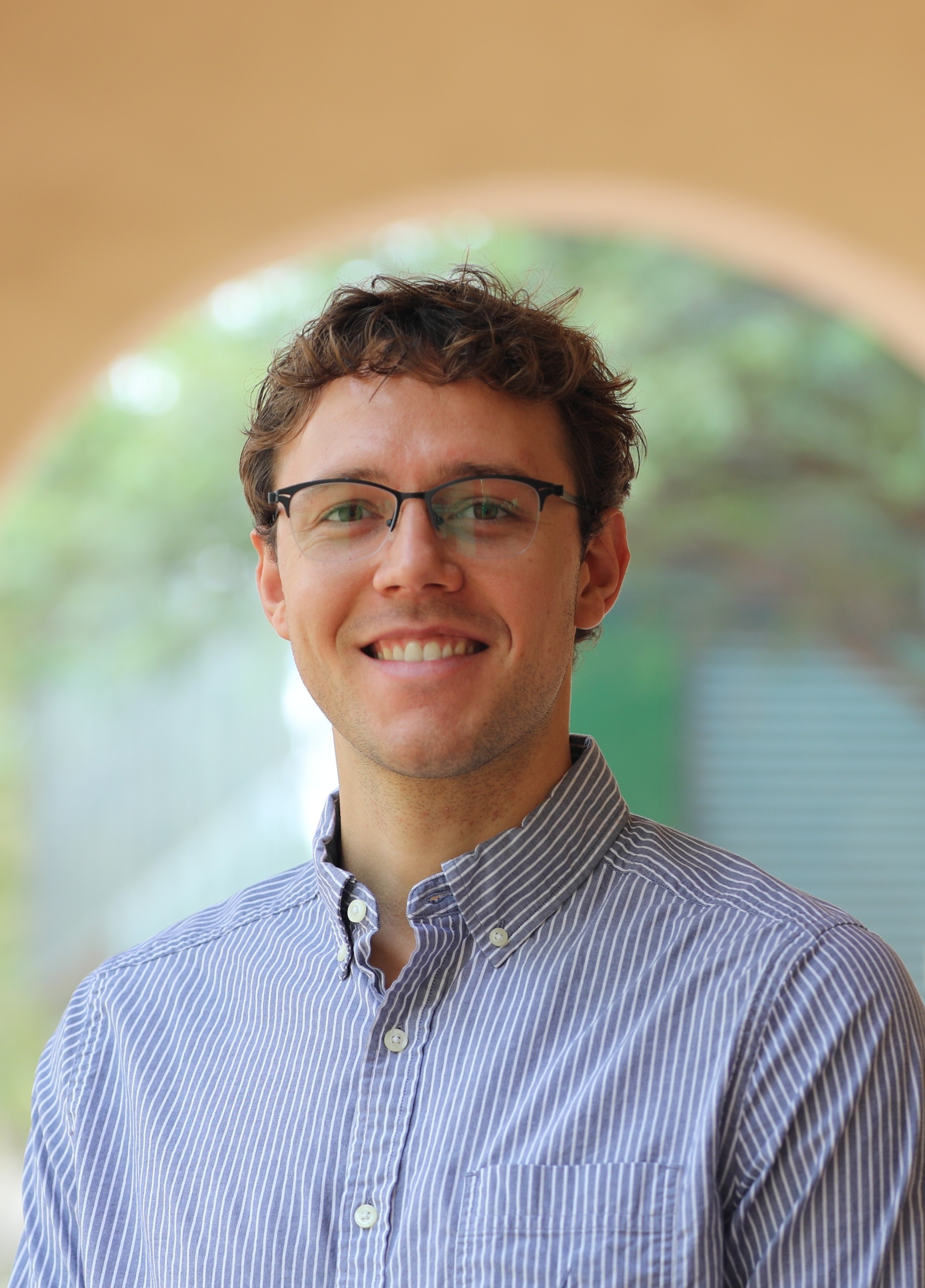}}]
{DAVID E. J. VAN WIJK}{\space} is a Postdoctoral Scholar in the Department of Mechanical and Civil Engineering at the California Institute of Technology. He received the Ph.D. degree in Aerospace Engineering from Texas A\&M University, College Station, TX, USA in 2025, and the B.S. degree in Mechanical Engineering from Cornell University, Ithaca, NY, USA, in 2021. His research interests include nonlinear dynamics and control, safety-critical control, and state estimation for robotics and autonomous systems.
\end{IEEEbiography}

\begin{IEEEbiography}
[{\includegraphics[width=1in,height=1.25in,clip,keepaspectratio]{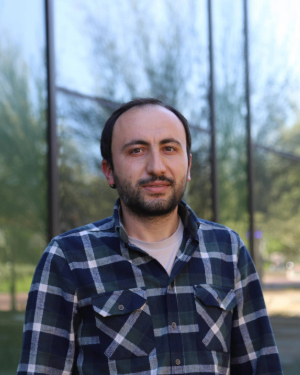}}]
{ERSIN DA\c{S}}{\space}(Member, IEEE) is an Assistant Professor in the Department of Mechanical, Materials, and Aerospace Engineering at Illinois Tech. He was previously a postdoctoral scholar at Caltech. He received an MSc in Mechatronics Engineering from Istanbul Technical University and his Ph.D. in Mechanical Engineering from Hacettepe University. His research interests include robust and safe control, state estimation, and machine learning for robotics and autonomous systems.
\end{IEEEbiography}

\begin{IEEEbiography}
[{\includegraphics[width=1in,height=1.25in,clip,keepaspectratio]{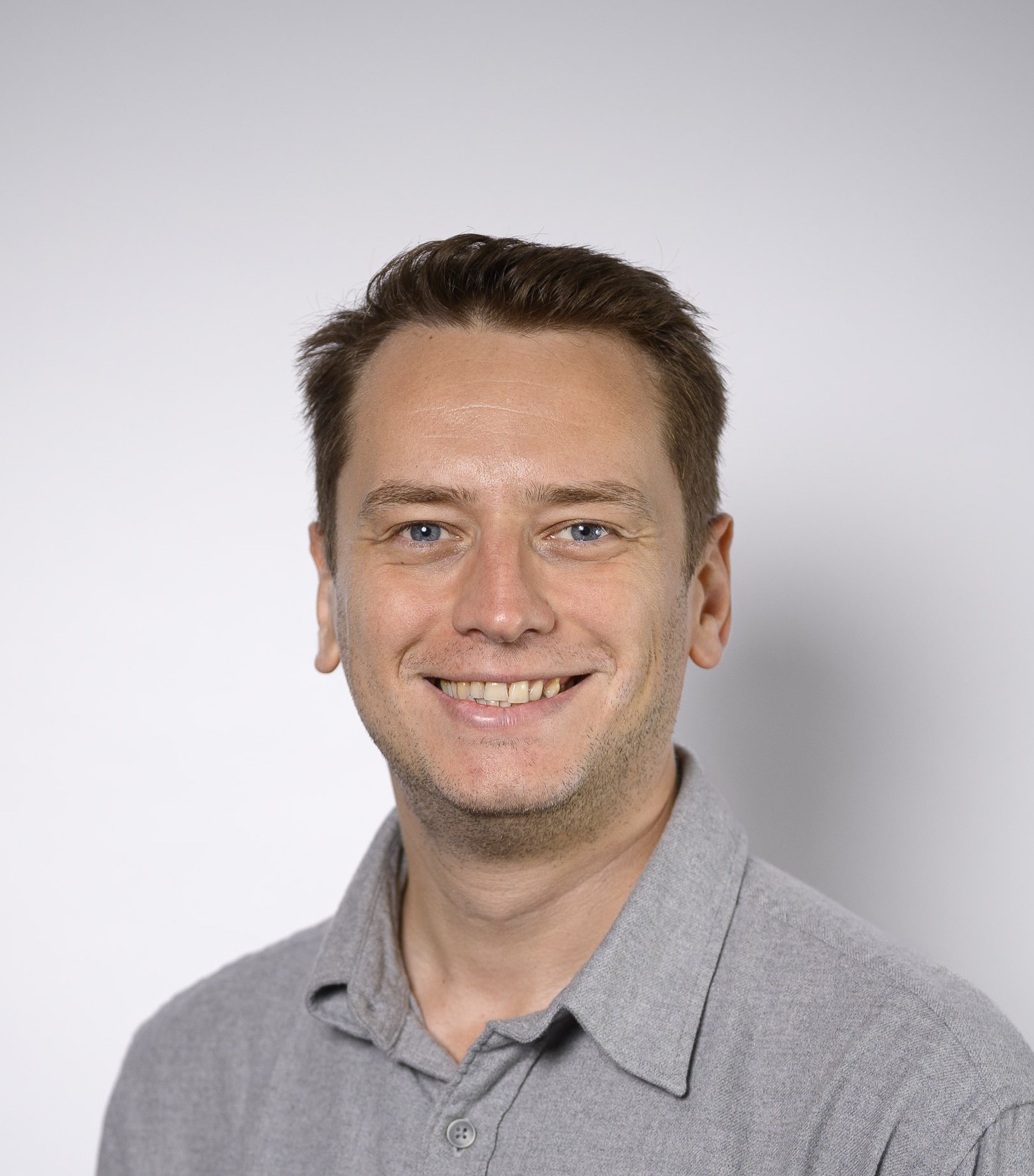}}]
{ANIL ALAN}{\space} completed his PhD in Mechanical Engineering at the University of Michigan, USA, where he was awarded the Rackham Predoctoral Fellowship and Professor Pierre T. Kabamba Award; and holds an MSc from Bilkent University, Turkey. Since 2026, he holds a Postdoc position at the Department of Electrical Engineering in TU Eindhoven, The Netherlands. Previously he worked at the Delft Center for Systems and control in TU Delft, The Netherlands as a Postdoc. His research focuses on safety-critical control certification for interconnected systems, with application to multi-agent networks such as intelligent vehicles.
\end{IEEEbiography}

\begin{IEEEbiography}
[{\includegraphics[width=1in,height=1.25in,clip,keepaspectratio]{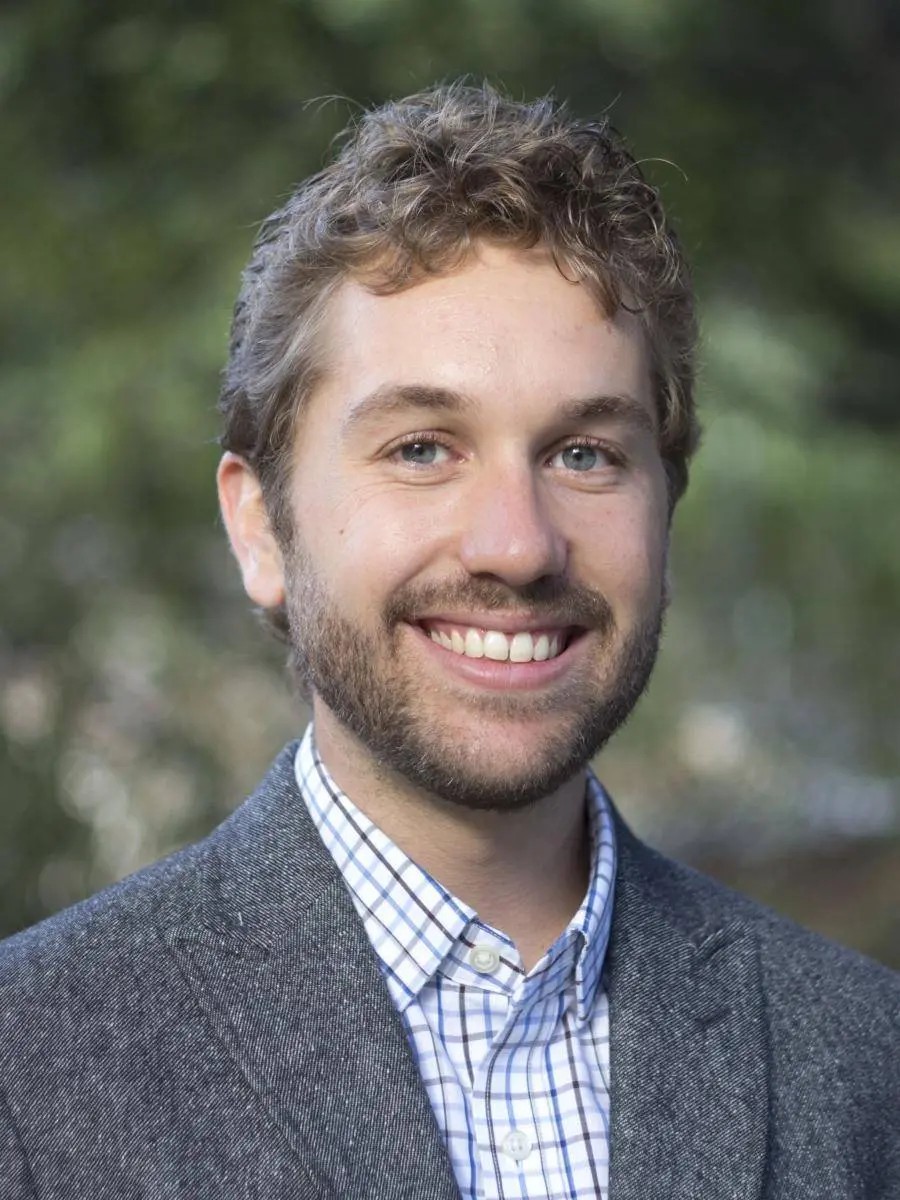}}]
{SAMUEL COOGAN}{\space}(Senior Member, IEEE) received the B.S. degree in electrical engineering from the Georgia Institute of Technology,
Atlanta, GA, USA, in 2010, and the M.S. and
Ph.D. degrees in electrical engineering from the
University of California, Berkeley, Berkeley, CA,
USA, in 2012 and 2015, respectively.
He is currently an Associate Professor and
the Demetrius T. Paris Junior Professor with the
School of Electrical and Computer Engineering
and the School of Civil and Environmental Engineering, Georgia Institute of Technology. Prior to joining Georgia Tech
in 2017, he was an Assistant Professor with the University of California,
Los Angeles, Los Angeles, CA, USA, from 2015 to 2017.
Dr. Coogan was the recipient of a career award from the National
Science Foundation in 2018, a Young Investigator Award from the Air
Force Office of Scientific Research in 2019, and the Donald P. Eckman
Award from the American Automatic Control Council in 2020. He is a
Member of SIAM.
\end{IEEEbiography}

\begin{IEEEbiography}
[{\includegraphics[width=1in,height=1.25in,clip,keepaspectratio]{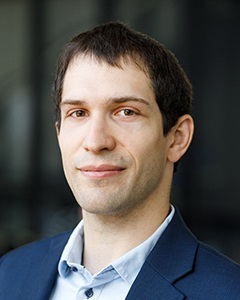}}]
{TAMAS G. MOLNAR}{\space}(Member, IEEE) is an Assistant Professor of Mechanical Engineering at the Cleveland State University since 2026. Beforehand, he was an assistant professor at the Wichita State University from 2023 to 2026. He held postdoctoral positions at the California Institute of Technology, from 2020 to 2023, and at the University of Michigan, Ann Arbor, from 2018 to 2020. He received the Ph.D. and M.S. degrees in Mechanical Engineering and the B.S. degree in Mechatronics Engineering from the Budapest University of Technology and Economics, Hungary, in 2018, 2015, and 2013. His research interests include nonlinear dynamics and control, safety-critical control, and time delay systems with applications to connected automated vehicles, robotic systems, and autonomous systems.
\end{IEEEbiography}

\begin{IEEEbiography}
[{\includegraphics[width=1in,height=1.25in,clip,keepaspectratio]{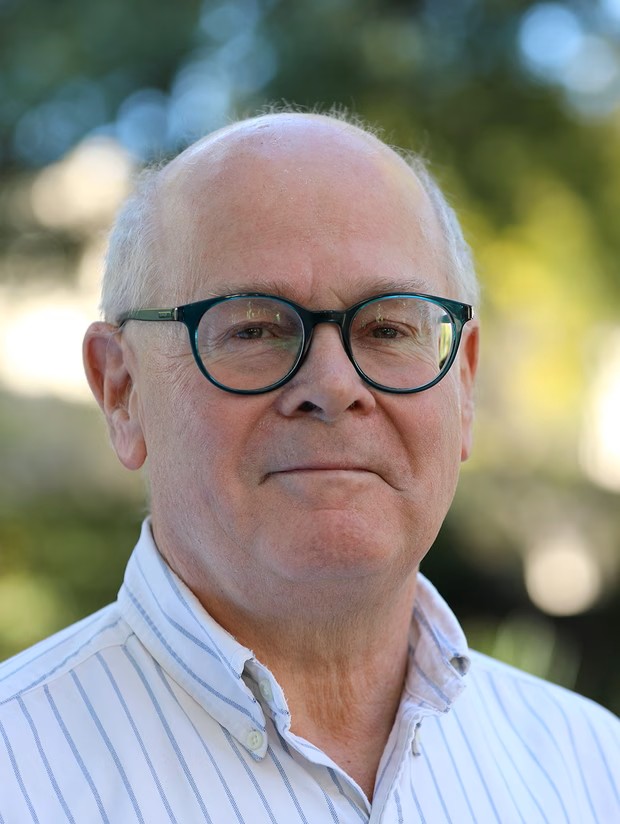}}]
{JOEL W. BURDICK}{\space}(Member, IEEE) the Richard L. and Dorothy M. Hayman Professor of Mechanical Engineering and Bioengineering, received his undergraduate degrees in mechanical
engineering and chemistry from Duke University and
M.S. and Ph.D. degrees in mechanical engineering from
Stanford University. He has been with the department of
mechanical engineering at the Caltech since May 1988,
where he has been the recipient of the NSF Presidential Young Investigator award, the Office of Naval Research
Young Investigator award, and the Feynman fellowship.
He has also received the ASCIT Award for Excellence in
Undergraduate Teaching and the GSA Award for Excellence in Graduate Student Education, and received the
Popular Mechanics Breakthrough Award in 2011. In addition to mechanical engineering, he is a core faculty of
the control and dynamical systems option, as well as a
faculty affiliate in the options of bioengineering (BE) and
computational and neural systems (CNS). His research
interests lie mainly in the areas of robotics, kinematics, and
mechanical systems.
\end{IEEEbiography}

\begin{IEEEbiography}[{\includegraphics[width=1in,height=1.25in,clip,keepaspectratio]{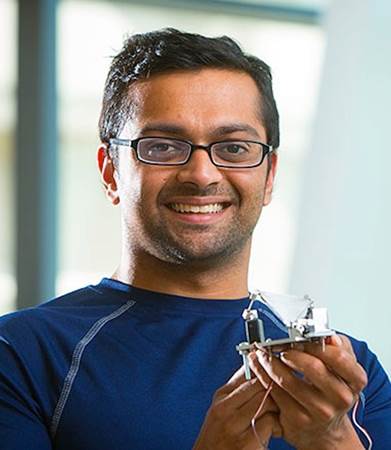}}]
{MANORANJAN MAJJI}{\space}(Senior Member, IEEE), received the B.E. (Hons) in mechanical engineering from the Birla Institute of Technology and Science, Pilani, India in 2003, and M.S. and Ph.D  in aerospace engineering from Texas A{\&}M University, College Station, in 2006 and 2009 respectively. 

After being a research associate working on computational vision at Texas A{\&}M, he was an Assistant Professor of Mechanical and Aerospace Engineering at the University at Buffalo, New York. He then joined the faculty of Texas A{\&}M University, where he is currently a Professor of Aerospace Engineering. He also directs the Land, Air and Space Robotics (LASR) laboratory. He has a diverse background in several aspects of dynamics and control of aerospace vehicles with expertise spanning the whole spectrum of modeling, analysis, computations and experiments. In the areas of state estimation, astrodynamics, tensegrity systems, vision navigation, and system identification, he has made fundamental contributions documented in over 150 publications. His 10 PhD graduates are making valuable contributions in the academic, national laboratory, and industrial research establishments. He is an associate editor of the Journal of Astronautical Sciences.

Dr. Majji is a senior member of the IEEE, an associate fellow of the American Institute of Aeronautics and Astronautics (AIAA) and a fellow of the American Astronautical Society (AAS). He is a recipient of the Milton Plesur award for excellence in teaching from the University at Buffalo, SUNY,  and was awarded the 2021 Dean of Engineering Excellence Award at Texas A{\&}M and is a 2021 Career Initiation Fellow of the Texas A{\&}M Institute of Data Science. In addition to being a scholar, Majji has a great deal of engineering experience developing software systems and embedded systems from OEM products. He holds a provisional patent on a simultaneous location and mapping software suite, and was awarded a patent for developing a novel omni-directional robot.
\end{IEEEbiography}

\begin{IEEEbiography}
[{\includegraphics[width=1in,height=1.25in,clip,keepaspectratio]{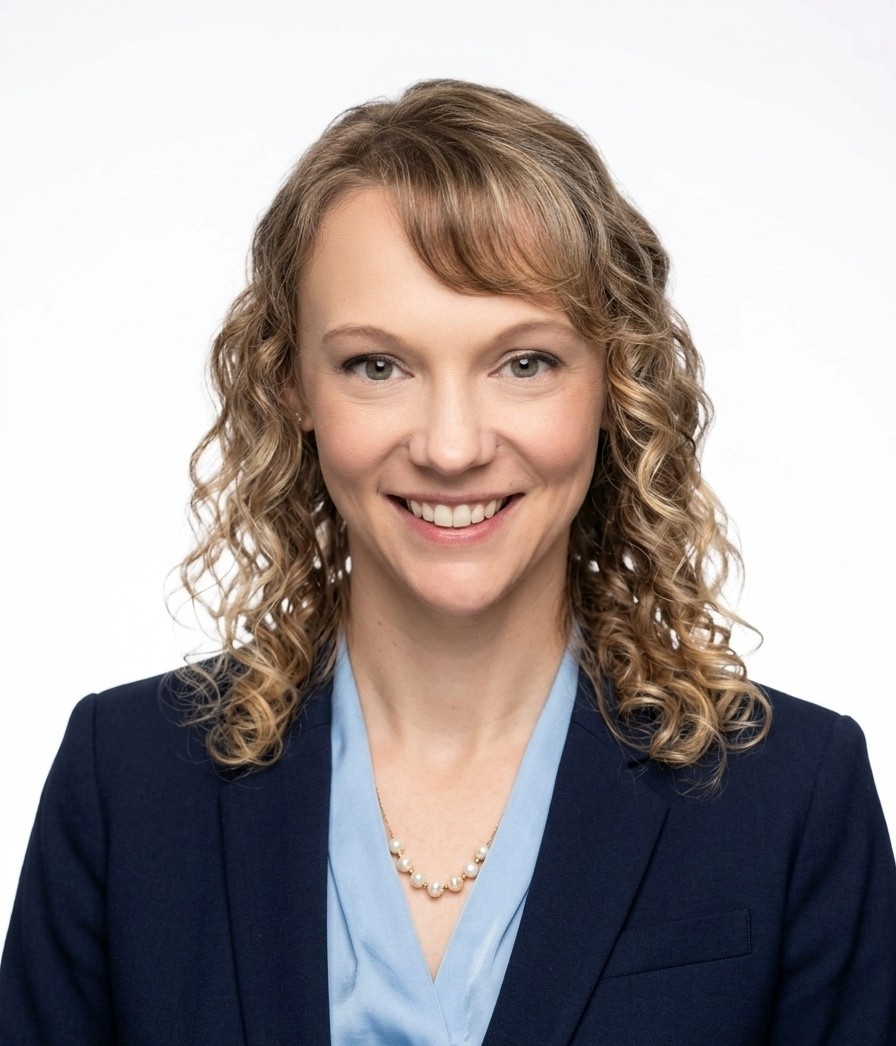}}]
{KERIANNE L. HOBBS}{\space}(Member, IEEE) 
leads space vehicle autonomy and trust research at The Aerospace Corporation. Previously, as Safe Autonomy Lead at the Air Force Research Laboratory, she directed the Safe Trusted Autonomy for Responsible Spacecraft (STARS) program, developing neural network control systems using reinforcement learning, runtime assurance technologies, and human-AI interfaces for multi-vehicle autonomous satellite operations. Her research expertise includes formal methods and runtime assurance for autonomous software assurance. Prior to STARS, she contributed to the development and testing of automatic collision avoidance technologies for the F-16. She holds a B.S. in Aerospace Engineering from Embry-Riddle Aeronautical University, an M.S. in Astronautical Engineering from the Air Force Institute of Technology, and a Ph.D. in Aerospace Engineering from Georgia Institute of Technology. Dr. Hobbs is a 2024 AIAA Associate Fellow and 2020 AFCEA 40 Under 40 recipient. She was part of the team awarded the 2018 Collier Trophy for the Automatic Ground Collision Avoidance System.
\end{IEEEbiography}

\end{document}

%% file: _aux/preamble.tex
\usepackage{amsthm}
\usepackage[colorlinks,urlcolor=blue,linkcolor=blue,citecolor=blue]{hyperref}
\usepackage{color,array}
\usepackage{graphicx}
\usepackage[T1]{fontenc}
\usepackage{subcaption}
\usepackage{amsmath} 
\usepackage{amssymb}
\usepackage{bm}
\usepackage{amsfonts}
\usepackage{dsfont}
\usepackage{cite}
\usepackage{cleveref}

\newtheorem{theorem}{Theorem}
\newtheorem{lemma}{Lemma}
\newtheorem{corollary}{Corollary}

\newtheorem{assumption}{Assumption}

\theoremstyle{definition}
\newtheorem{definition}{Definition}
\newtheorem{example}{Example}
\newtheorem{casestudy}{Case Study}

\theoremstyle{remark}
\newtheorem{remark}{Remark}

\makeatletter
\def\@IEEEproof[#1]{\par\noindent{{\itshape #1:\ }}\ignorespaces}
\makeatother

\newcommand{\longthmtitle}[1]{\mbox{}{\textit{(#1):}}}

\newcommand{\R}{\mathbb{R}}

\newcommand{\eye}{\boldsymbol{I}}

\newcommand{\x}{\boldsymbol{x}}

\newcommand{\sspace}{\mkern9mu}
\newcommand{\nspace}[1]{\mkern#1mu}

\newcommand{\norm}[1]{\left\Vert #1 \right\Vert}

\newcommand{\Cs}{\mathcal{C}_{\rm S}} 
\newcommand{\Cb}{\mathcal{C}_{\rm B}} 
\newcommand{\Cbi}{\mathcal{C}_{\rm I}} 
\newcommand{\Cd}{\mathcal{C}_{\rm D}(t)}
\newcommand{\Cdhat}{\widehat{\mathcal{C}}_{\rm D}(t)}

\newcommand{\Cdcon}{\mathcal{C}_{\rm D}}
\newcommand{\Cdhatcon}{\widehat{\mathcal{C}}_{\rm D}}

\newcommand{\Tt}{T} 

\newcommand{\ub}{\boldsymbol{k}_{\rm b}} 

\newcommand{\phidg}[2]{\boldsymbol{\phi}^{d} (#1, #2)}
\newcommand{\phidb}[2]{\boldsymbol{\phi}^{d}_{\rm b} (#1, #2)}
\newcommand{\phidbhat}[2]{\boldsymbol{\phi}^{\hat{d}}_{\rm b} (#1, #2)}

\newcommand{\phinb}[2]{\boldsymbol{\phi}_{\rm b} (#1, #2)}

\newcommand{\phinom}{\phinb{\tau}{\boldsymbol{x}}}
\newcommand{\phinomT}{\phinb{\Tt}{\boldsymbol{x}}}

\newcommand{\phidhat}{\phidbhat{\tau}{\boldsymbol{x}}}
\newcommand{\phidhatT}{\phidbhat{\Tt}{\boldsymbol{x}}}

\newcommand{\stmnom}{\boldsymbol{\Phi}_{\rm b}(\tau,\boldsymbol{x})}
\newcommand{\stmnomT}{\boldsymbol{\Phi}_{\rm b}(\Tt,\boldsymbol{x})}
\newcommand{\jac}{\boldsymbol{J}_{\rm cl}}

\newcommand{\stmdhat}{\boldsymbol{\Phi}(\tau,\boldsymbol{x})}
\newcommand{\stmdhatT}{\boldsymbol{\Phi}(\Tt,\boldsymbol{x})}

\newcommand{\phid}{\phidb{\tau}{\boldsymbol{x}}}

\newcommand{\phidT}{\phidb{\Tt}{\boldsymbol{x}}}

\newcommand{\tb}{\tau}

\newcommand{\xzero}{\boldsymbol{x}_0}

\definecolor{darkblue}{RGB}{0,0,102}
\definecolor{lightblue}{RGB}{77,77,148}

\definecolor{gold}{RGB}{234, 170, 0}
\definecolor{metallic_gold}{RGB}{139, 111, 78}

\newcommand{\bxi}{\bm{\xi}}

\newcommand{\dhat}{\hat{\boldsymbol{d}}}
\newcommand{\bLam}{\boldsymbol{\Lambda}}
\newcommand{\dmax}{\delta_{\rm max}}
\newcommand{\dhatdot}{\skew{7}{\dot}{\skew{7}{\hat}{\bm{d}}}}

%% file: sections/01_Abstract.tex
\begin{abstract}
Designing safe controllers is crucial and notoriously challenging for input-constrained safety-critical control systems.
Backup control barrier functions offer an approach for the construction of safe controllers online by considering the flow of the system under a backup controller. However, in the presence of model uncertainties, the flow cannot be accurately computed, making this method insufficient for safety assurance. To tackle this shortcoming, we integrate backup control barrier functions with uncertainty estimators and calculate the flow under a reconstruction of the model uncertainty while refining this estimate over time.
We prove that the controllers resulting from the proposed \textit{Uncertainty Estimator Backup Control Barrier Function (UE-bCBF)} approach guarantee safety, are robust to unknown disturbances, and satisfy input constraints.
\end{abstract}

%% file: sections/02_Introduction.tex
\section{Introduction}

Controllers that satisfy safety constraints are of paramount importance for many autonomous systems. \textit{Control barrier functions (CBFs)} \cite{ames_2017} offer a simple and effective approach for safety-critical control by providing sufficient conditions for forward invariance of safe sets. However, designing such safe sets for input-constrained systems remains a challenge, especially for high-dimensional dynamics and complex state constraints. Furthermore, CBFs rely on an accurate model of the system dynamics, but such models are rarely without errors in real-world applications. In this paper, we seek to address both of these challenges concurrently. 

The safety-critical control literature is rich with attempts to accommodate model mismatches. Robust methods \cite{jankovic_robust_2018,garg_robust_2021,shaw_cortez_control_2021} address disturbances typically through worst-case analysis that provides safety guarantees using an additional robustifying term. Input-to-state safety \cite{issf_og} addresses input disturbances \cite{Issf_ames} and can be made less conservative via a tunable robustness parameter \cite{alan2022tunableissf}.
Uncertainty estimators such as disturbance observers have been used to reconstruct a representation of the disturbance signal
\cite{das2025robustcontrolbarrierfunctions,wang2023DOB,Isaly2024DOB}. 
Adaptive methods,
which are effective in handling parametric uncertainty, can also reduce conservatism \cite{lopez_adaptive_robustCBF}. Data-driven
\cite{emam_data_driven}
and learning-based
\cite{lindemann2024learningrobustoutputcontrol}
approaches show promise in handling uncertainty in dynamics or states for real-world systems. While these approaches present viable solutions to addressing model uncertainties, they assume that such a safe set can be found explicitly that will lead to the satisfaction of input constraints---a strong assumption for most systems.

To design safe sets in which every state has a safe control action (i.e., controlled invariant sets), we leverage \textit{backup control barrier functions (bCBFs)} \cite{gurriet_online_2018,gurriet_scalable_2020} which construct controlled invariant sets online by examining the predicted state evolution under a backup control policy. While this approach guarantees safety with input constraints, it is sensitive to model uncertainties because it requires forward integrating the uncertain model.
To remedy this, our previous work~\cite{vanWijk_DRbCBF_24} derived conditions for online controlled invariance in the presence of disturbances.
We used an upper bound on the state evolution uncertainty through a worst-case analysis, resulting in conservative safety constraints in some cases. 

In this work, we introduce an approach to online controlled invariance in the presence of disturbances, and reduce conservatism via uncertainty estimators.
Our main contribution is \textit{uncertainty estimator backup CBFs}---a novel class of CBFs for the safety-critical control of input-constrained uncertain systems.
The proposed method uses state predictions under the reconstruction of the disturbance to define a subset inside a controlled invariant safe set of the disturbed system.
We derive forward invariance conditions for such a subset, and we show that these are made less conservative over time using the 
uncertainty estimator.
We use these conditions to design robust safety-critical controllers, which account for the evolution of the uncertainty estimator error and for the sensitivity of state predictions to the estimated disturbance, instead of merely adding an uncertainty estimator to existing methods like~\cite{vanWijk_DRbCBF_24}. This generalizes the proposed approach to be utilized across a large range of possible uncertainty estimators and predictors.
We prove that the proposed approach guarantees safety for systems with limited control authority even in the presence of unknown, bounded disturbances.

The remainder of the paper is organized as follows. Section~\ref{sec:prelim} reviews CBFs and bCBFs.
In Section~\ref{sec:main_res}, we formalize uncertainty estimators in the context of implicit set safety, and present the proposed
uncertainty estimator backup CBF approach.
In Section~\ref{sec:examples}, we cover the results of our numerical simulations, and we conclude with Section~\ref{sec:conclusion}.

%% file: sections/03_Preliminaries.tex
\section{Preliminaries} \label{sec:prelim}

\subsection{Control Barrier Functions} \label{sec:CBF}
Consider a nonlinear control affine system of the form
\begin{align} \label{eq:affine-dynamics}
    \dot{\boldsymbol{x}} = \boldsymbol{f}(\boldsymbol{x}) + \boldsymbol{g}(\boldsymbol{x})\boldsymbol{u}, \nspace{8}
    \boldsymbol{x} \in \mathcal{X} \subset \mathbb{R}^n, \nspace{8}
    \boldsymbol{u} \in \mathcal{U} \subseteq \mathbb{R}^m,
\end{align}
with continuously differentiable functions ${\boldsymbol{f}\!:\!\mathcal{X} \!\rightarrow\! \mathbb{R}^n}$ and ${\boldsymbol{g}\!:\!\mathcal{X} \!\rightarrow \!\mathbb{R}^{n \times m}}$.
We assume that the admissible input set $\mathcal{U}$ is an $m$-dimensional convex polytope.
For an initial condition ${\boldsymbol{x}_0 \triangleq \boldsymbol{x}(0)  \in \mathcal{X}}$, if $\boldsymbol{u}$ is given by a locally Lipschitz controller ${\boldsymbol{k}:\mathcal{X} \rightarrow \mathcal{U}}$, ${\boldsymbol{u}=\boldsymbol{k}(\boldsymbol{x})}$, the closed-loop system has a unique solution.

A notion of safety is defined by membership to a set $\Cs$, and safe controllers render this safe set forward invariant.
A set ${\mathcal{C} \subset \mathbb{R}^n}$ is \textit{forward invariant} along the closed-loop system if ${\boldsymbol{x}(0) \in \mathcal{C} \!\implies\! \boldsymbol{x}(t) \in \mathcal{C},}$ for all ${t > 0}$.
Consider the safe set $\Cs$ as the 0-superlevel set of a continuously differentiable function ${h : \mathcal{X} \rightarrow \mathbb{R}}$ with
${\Cs \triangleq \{\boldsymbol{x} \in \mathcal{X} : h(\boldsymbol{x}) \ge 0\}}$,
where the gradient of $h$ along the boundary of $\Cs$ remains nonzero.
A function ${h : \mathcal{X} \rightarrow \mathbb{R}}$ is a CBF \cite{ames_2017} for \eqref{eq:affine-dynamics} on $\Cs$ if there exists a class-$\mathcal{K}_{\infty}$ function\footnote{The function ${\alpha : \mathbb{R}_{\ge 0} \rightarrow \mathbb{R}_{\ge 0}}$ is of class-$\mathcal{K}_{\infty}$ if it is continuous, strictly increasing, $\alpha(0)=0$, and $\text{lim}_{x \rightarrow \infty} \nspace{2}\alpha(x) = \infty$.} $\alpha$ such that for all ${\boldsymbol{x} \in \Cs}$
\begin{equation*}
    \sup_{\boldsymbol{u} \in \mathcal{U}} \dot{h}(\boldsymbol{x},\boldsymbol{u}) =
    \nabla h(\boldsymbol{x}) \big( \boldsymbol{f}(\boldsymbol{x}) + \boldsymbol{g}(\boldsymbol{x}) \boldsymbol{u} \big)
    >  -\alpha(h(\boldsymbol{x})).
\end{equation*}
\begin{theorem}
[\hspace{-0.01cm}\cite{ames_2017}]
\label{thm: cbf}
If $h$ is a CBF for \eqref{eq:affine-dynamics} on $\Cs$, then any locally Lipschitz controller ${\boldsymbol{k}:\mathcal{X} \rightarrow \mathcal{U}}$, ${\boldsymbol{u}=\boldsymbol{k}(\boldsymbol{x})}$ satisfying 
\begin{align} \label{eq: cbf_condition}
    \nabla h(\boldsymbol{x}) \big( \boldsymbol{f}(\boldsymbol{x}) + \boldsymbol{g}(\boldsymbol{x}) \boldsymbol{u} \big) \ge -\alpha(h(\boldsymbol{x})),
\end{align}
for all ${\boldsymbol{x} \in \Cs}$ renders the set $\Cs$ forward invariant.
\end{theorem}

One can ensure the safety of \eqref{eq:affine-dynamics} for a primary controller
${\boldsymbol{k}_{\rm p}\!:\!\mathcal{X}\!\to\!\nspace{2}\mathcal{U}}$
by solving the following quadratic program (QP):
\begin{align} 
    \boldsymbol{k}_{\rm safe}(\boldsymbol{x}) = \underset{\boldsymbol{u} \in \mathcal{U}}{\text{arg\,min}} \mkern9mu &
    \left\Vert \boldsymbol{k}_{\rm p}(\x)-\boldsymbol{u}\right\Vert^{2}
    \tag{CBF-QP} \label{eq:cbf-qp} \\
    \text{s.t.} \quad &
    \nabla h(\boldsymbol{x}) \big( \boldsymbol{f}(\boldsymbol{x}) + \boldsymbol{g}(\boldsymbol{x}) \boldsymbol{u} \big) \ge -\alpha(h(\boldsymbol{x})).
    \nonumber
\end{align}
Ensuring the feasibility of the \eqref{eq:cbf-qp} can be challenging, especially for high-dimensional systems with input bounds. This motivates a related tool known as backup CBFs.

\subsection{Backup Control Barrier Functions} \label{sec:bCBF}

Backup CBFs \cite{gurriet_online_2018,gurriet_scalable_2020} construct controlled invariant sets online for feasibility guarantees with input bounds. A set ${\mathcal{C} \subset \mathbb{R}^n}$ is \textit{controlled invariant} if there exists a controller ${\boldsymbol{k}:\mathcal{X} \rightarrow \mathcal{U}}$, ${\boldsymbol{u}=\boldsymbol{k}(\boldsymbol{x})}$ rendering $\mathcal{C}$ forward invariant for \eqref{eq:affine-dynamics}. 

As in Section~\ref{sec:CBF}, assume that safety is defined by a set $\Cs$ which is not necessarily controlled invariant.
Now, suppose there exists a backup set and corresponding backup controller with the following properties.
\begin{definition}\label{def: backup} A set ${\Cb \! \triangleq \! \{\boldsymbol{x} \!\in\! \mathcal{X} \!:\! h_{\rm b}(\boldsymbol{x}) \!\ge\! 0\}}$ is a {backup set} if ${\Cb \subseteq \Cs}$ and $\Cb$ is controlled invariant. A continuously differentiable control law ${\boldsymbol{k}_{\rm b}: \mathcal{X} \rightarrow \mathcal{U}}$ rendering $\Cb$ forward invariant is called a {backup controller}.
\end{definition}
For example, one can define a backup set by a level set of a control Lyapunov function centered on a stabilizable equilibrium point for the linearized dynamics, and then render this set forward invariant with a simple feedback law
\cite{gurriet_scalable_2020}.
The closed-loop system under ${\boldsymbol{k}_{\rm b}}$ is denoted as
\begin{align}\label{eq: f_cl}
    \dot{\boldsymbol{x}} = \boldsymbol{f}_{\rm cl}(\boldsymbol{x}) \triangleq \boldsymbol{f}(\boldsymbol{x}) + \boldsymbol{g}(\boldsymbol{x})\ub(\boldsymbol{x}).
\end{align}

Backup control barrier functions construct an implicit representation of a new safe set by forward integrating the dynamical system over a finite time horizon. By defining the implicit safe set as all states which can safely reach the backup set using the backup controller, recursive feasibility and safety is guaranteed thanks to Definition~\ref{def: backup}. To be more precise, this implicit safe set ${\Cbi \!\subseteq\! \Cs}$ is defined as
\begin{align}
    \Cbi \triangleq \left\{ \boldsymbol{x} \in \mathcal{X} \,\middle|\, 
    \begin{array}{c}
    h(\phinom) \geq 0, \forall \nspace{1} \tau \in [0,T], \\
    h_{\rm b}(\phinomT) \geq 0 \\
    \end{array}
    \right\},
\end{align}
where $\phinom$ is the \textit{flow} of the backup system~\eqref{eq: f_cl} over the interval ${\tau \!\in\! [0,T]}$ for a horizon ${T \!>\! 0}$ starting at state $\boldsymbol{x}$:
\begin{align} \label{eq: nomFlow}
    \frac{\partial}{\partial \tau}{\boldsymbol{\phi}_{\rm b}}(\tau,\boldsymbol{x}) = \boldsymbol{f}_{\rm cl}(\phinom), \sspace \phinb{0}{\boldsymbol{x}} = \boldsymbol{x}.
\end{align}
A controller ${\boldsymbol{k}:\mathcal{X} \rightarrow \mathcal{U}}$, ${\boldsymbol{u}=\boldsymbol{k}(\boldsymbol{x})}$ makes $\Cbi$ forward invariant, and thus~\eqref{eq:affine-dynamics} safe w.r.t.~$\Cs$, if there exist class-$\mathcal{K}_{\infty}$ functions $\alpha$, $\alpha_{\rm b}$ such that
\begin{subequations}\label{eq: nom_bcs1}
\begin{align} 
    \nabla h(\phinom)\stmnom\dot{\boldsymbol{x}} &\ge - \alpha (h(\phinom)), \label{eq: htraj_nom}  \\ 
    \!\!\nabla h_{\rm b}(\phinomT) \stmnomT \dot{\boldsymbol{x}} &\ge - \alpha_{\rm b} (h_{\rm b}(\phinomT)), \label{eq: hb_nom}
\end{align}
\end{subequations}
for all ${\tau \in [0,T]}$ and ${\x\in\mathcal{\Cbi}}$. Here, ${\dot{\boldsymbol{x}}= \boldsymbol{f}(\boldsymbol{x}) + \boldsymbol{g}(\boldsymbol{x})\boldsymbol{u}}$, and ${\stmnom \triangleq \partial \phinom/\partial \boldsymbol{x}}$ is the state-transition matrix capturing the sensitivity of the flow to perturbations in the initial state $\boldsymbol{x}$. The state-transition matrix is the solution to
\begin{equation} \label{eq: stm_nominal}
  \begin{gathered} 
    \frac{\partial}{\partial \tau}{{\boldsymbol{\Phi}}}_{\rm b}(\tau, \boldsymbol{x}) \!=\! \jac(\phinom)\stmnom,
    \sspace
    \boldsymbol{\Phi}_{\rm b}(0,\boldsymbol{x}) = \boldsymbol{I},
\end{gathered}
\end{equation}
where $\boldsymbol{I}$ is the ${n \!\times\! n}$ identity matrix, and
\begin{align} \label{eq:F_cl_def}
    \jac(\boldsymbol{x})\triangleq \frac{\partial \boldsymbol{f}_{\rm cl}(\boldsymbol{x})}{\partial \boldsymbol{x}},
\end{align}
is the Jacobian of $\boldsymbol{f}_{\rm cl}$ in~\eqref{eq: f_cl}, that is evaluated at $\phinom$.

Because \eqref{eq: htraj_nom} must hold for all $\tau$ in an uncountable set, namely, $[0,T]$, this constraint
is typically enforced for a discrete selection of times in $[0,T]$.
Then, the safety of a primary controller can be enforced similar to the \eqref{eq:cbf-qp}:
\begin{align*} 
    \boldsymbol{k}_{\rm safe}(\boldsymbol{x}) = \underset{\boldsymbol{u} \in \mathcal{U}}{\text{arg\,min}} \mkern9mu &
    \left\Vert \boldsymbol{k}_{\rm p}(\x)-\boldsymbol{u}\right\Vert^{2} \quad \tag{bCBF-QP} \label{eq:bcbf-qp} \\
    \text{s.t.  } 
    & \eqref{eq: htraj_nom}, \ \eqref{eq: hb_nom},
\end{align*}
for all ${\tau \in \{0, \Delta, \dots, \Tt \}}$ where ${\Delta > 0}$ is a discretization time step satisfying ${T/\Delta \in \mathbb{N}}$.
Unlike the \eqref{eq:cbf-qp}, if the backup controller is designed such that ${\ub(\boldsymbol{x}) \in \mathcal{U}}$ for all ${\boldsymbol{x} \in \Cbi}$ then the feasibility of the~\eqref{eq:bcbf-qp} is guaranteed over $\Cbi$ for appropriate $\alpha$ and $\alpha_{\rm b}$ \cite{gurriet_scalable_2020}. Upon performing this discretization, it has been shown that one can robustify the condition \eqref{eq: htraj_nom} between sample times \cite[Thm.~1]{gurriet_scalable_2020}.

%% file: sections/04_MainResult.tex
\section{Main Results} \label{sec:main_res}
While the backup CBF approach ensures safety for input-constrained systems, it implicitly assumes that the dynamical model is perfect. In practice, external or internal disturbances may cause the evolution of the state to be uncertain. We seek to use the advantages of backup CBFs even in the presence of unknown disturbances.
Consider a nonlinear affine system
\begin{align} \label{eq: disturbed_dyn}
    \dot{\boldsymbol{x}} = \boldsymbol{f}(\boldsymbol{x}) + \boldsymbol{g}(\boldsymbol{x})\boldsymbol{u} + \boldsymbol{d}(t),
\end{align}
where ${\boldsymbol{d}(t) \in \mathcal{D} \subset \mathbb{R}^n}$ is a continuously differentiable unknown additive process disturbance. For an initial condition ${\boldsymbol{x}(t_0) = \boldsymbol{x}_0 \in \mathcal{X}}$ and a locally Lipschitz controller ${\boldsymbol{u}=\boldsymbol{k}(\boldsymbol{x})}$, the closed-loop system has a unique solution $\phidg{t}{\xzero}$ over an interval of existence.
For the rest of the manuscript, we take ${t_0 = 0}$ without loss of generality.
We make the following assumption on the disturbance.
\begin{assumption} \label{as:disturbance}
There exists a known ${\delta_d > 0}$ and ${\delta_v > 0}$ such that ${\norm{\boldsymbol{d}(t)} \leq \delta_d}$ and ${\| \dot{\boldsymbol{d}}(t) \| \leq \delta_v}$  for all ${t \geq 0}$.
\end{assumption}
\begin{remark}
   Assumption~\ref{as:disturbance} is reasonable for many systems (e.g., \cite{DU2016207,9640564,yan2023surviving}), as some knowledge of the disturbance and its rate of change can be expected. In practice, the bounds $\delta_d, \delta_v$ can be computed from simulation or test data. 
\end{remark}
\subsection{Uncertainty Estimators}
The work in \cite{vanWijk_DRbCBF_24} addresses online controlled invariance for disturbed dynamical systems in the form \eqref{eq: disturbed_dyn} by computing a nominal flow which ignores the disturbance, and robustifies against potential flow errors caused by the disturbances. While effective in some scenarios, the method can be conservative as it robustifies against the worst-case disturbance at all times. To reduce conservatism, researchers often utilize techniques to obtain an estimate of the disturbance or uncertainty over time by filtering the discrepancy between the
measured states and the nominal system dynamics.
The reader is referred to \cite{UE_survey_2016} for a comprehensive survey on related techniques.
Building on this idea, we consider an \textit{error-quantified uncertainty estimator} of the form
\begin{subequations}\label{eq:uncertainty_estimator}
\begin{align}
    \hat{\boldsymbol{d}} &= \boldsymbol{q}(t,\x,\boldsymbol{\xi}), \\
    \dot{\boldsymbol{\xi}} &=  \boldsymbol{p}(t,\x,\boldsymbol{u},\boldsymbol{\xi}),
\end{align}   
\end{subequations}
where ${\hat{\boldsymbol{d}} \nspace{-1}\in\nspace{-1} \mathbb{R}^n}$ is a time-varying estimate of the uncertainty in the dynamics, ${\boldsymbol{\xi} \nspace{-1} \in \nspace{-1} \mathbb{R}^{n}}$ is an auxiliary state, and the continuously differentiable functions $\boldsymbol{p}, \boldsymbol{q} \nspace{-1} : \nspace{-1} \mathbb{R}_{\geq 0} \nspace{-1}\times\nspace{-1} \mathcal{X}\nspace{-1} \times\nspace{-1} \mathcal{U}\nspace{-1} \times \nspace{-1}\mathbb{R}^n \nspace{-1} \rightarrow  \nspace{-1} \mathbb{R}^n$ describe the evolution of the uncertainty estimate. An error-quantified uncertainty estimator is defined below.
\begin{definition}\longthmtitle{Error-quantified Uncertainty Estimator} \label{def:EQUE}
    The uncertainty estimator in \eqref{eq:uncertainty_estimator} is called an error-quantified uncertainty estimator for \eqref{eq: disturbed_dyn} if there exists a known time-varying bound $\Bar{e}(t)$ on the uncertainty error $\boldsymbol{e}(t) \triangleq \boldsymbol{d}(t) - \hat{\boldsymbol{d}}(t)$ such that $\norm{\boldsymbol{e}(t)} \leq \Bar{e}(t)$ for all ${t \geq 0}$. 
\end{definition}
For reasons that will become clear in the subsequent sections, we may choose any uncertainty estimator satisfying Definition~\ref{def:EQUE} to estimate the disturbance.

\begin{example}\longthmtitle{Disturbance Observer} \label{ex: DOB}
One possible form of an error-quantified uncertainty estimator is a simple first-order disturbance observer from \cite{das2025robustcontrolbarrierfunctions} which is given by
\begin{subequations} \label{eq:DOB_all}
\begin{align} \label{eq:DOB1}
    \dhat &= \boldsymbol{\Lambda} \big( \x - \boldsymbol{\xi} \big), \\
    \dot{\bxi} &= {\boldsymbol{f}}(\x) + {\boldsymbol{g}}(\x) \boldsymbol{u} + \dhat,  \label{eq:DOB2}
\end{align}
\end{subequations}
where ${\dhat \nspace{-1} \in \mathbb{R}^{n}}$ is the estimated disturbance, ${\bxi \nspace{-1} \in \mathbb{R}^{n}}$ is the auxiliary state, and ${\bLam \in \mathbb{R}^{n \times n}}$ is a diagonal positive definite gain matrix. Without a priori information, we can set ${\dhat(0) = {\bf 0}}$ by assigning ${\bxi(0) = \x_{0}}$. Note that the disturbance observer in \eqref{eq:DOB1}, \eqref{eq:DOB2} implicitly assumes that the state is perfectly known (similar to many other uncertainty estimators).
The following lemma gives a bound on the disturbance estimation error, ${\boldsymbol{e}(t)}$, thereby satisfying Definition~\ref{def:EQUE} and
indicating that this form of uncertainty estimator is appropriate for safe control design.
\end{example}

\begin{lemma}\label{lemma: dob_errorBound}
    For the disturbance observer \eqref{eq:DOB1}, \eqref{eq:DOB2} and system \eqref{eq: disturbed_dyn} satisfying Assumption~\ref{as:disturbance}, the disturbance estimation error ${\boldsymbol{e}(t) = \boldsymbol{d}(t) - \dhat(t)}$ is bounded by:
    \begin{align}
      \norm{\boldsymbol{e}(t)}
      \le
      {\rm e}^{-\lambda_{\min} t} \,\delta_d
      +
      \frac{\delta_v}{\lambda_{\min}}
      \bigl(
        1 - {\rm e}^{-\lambda_{\min} t}
      \bigr) \triangleq \bar{e}(t),
      \label{eq:ebar}
    \end{align}
    where $\lambda_{\min}$ is the minimum eigenvalue of $\boldsymbol{\Lambda}$.
    \begin{proof}
        Expressing the error dynamics, ${\dot{\boldsymbol{e}}(t) = \dot{\boldsymbol{d}}(t) - \boldsymbol{\Lambda}\boldsymbol{e}(t)}$,
        and integrating from $0$ to $t$ results in 
        \begin{align*} 
            \boldsymbol{e}(t) = {\rm e}^{-\bLam t} \boldsymbol{e}(0) + \int_{0}^{t} {\rm e}^{-\bLam(t-\vartheta)}\dot{\boldsymbol{d}}(\vartheta) \nspace{3} d\vartheta.
        \end{align*}
        Taking the norm, with ${\| \dot{\boldsymbol{d}}(t) \| \leq \delta_v}$ and ${\|{\rm e}^{-\bLam t}\| \le {\rm e}^{-\lambda_{\min} t}}$ because $\bLam$ is diagonal and positive definite, we have
        \begin{align*} 
            \norm{\boldsymbol{e}(t)} \leq {\rm e}^{-\lambda_{\rm min} t}\norm{\boldsymbol{e}(0)} + \int_{0}^{t} {\rm e}^{-\lambda_{\rm min} (t - \vartheta)}\delta_v \nspace{3} d\vartheta.
        \end{align*}
        Integrating and noticing that ${\norm{\boldsymbol{e}(0)} \leq \delta_d}$ yields~\eqref{eq:ebar}.
    \end{proof}
\end{lemma}
With the error bound established, we proceed to derive conditions for online controlled invariance with disturbances.

\subsection{Safety Conditions}

To establish safety in the presence of the disturbance, consider a backup set $\Cb$ and a backup controller $\ub$.
Assume now that $\ub$ makes $\Cb$ \textit{robustly} forward invariant, which is made more precise below.
\begin{assumption} \label{assump: robust_inv}
     The backup controller $\boldsymbol{k}_{\rm b}$ renders the backup set $\Cb$ forward invariant along \eqref{eq: disturbed_dyn} for any disturbance $\boldsymbol{d}(t)$ which satisfies $\norm{\boldsymbol{d}(t)} \leq \delta_d$ for all $t \geq 0$.
\end{assumption}
Such robustly forward invariant backup sets can be obtained, for example, by robustifying the level sets of Lyapunov functions, which has been studied extensively in the literature \cite{jankovic_robust_2018},
\cite[Ch. 13.1]{khalil2002nonlinear}. While the robustification will likely result in a smaller backup set, with our approach, the smaller backup set is then expanded implicitly to generate a larger controlled invariant set online.

Given a robust backup controller, consider two separate flows: the flow under the true disturbance, denoted $\boldsymbol{\phi}^{d}_{\rm b}(\tau,\x)$:
\begin{align} \label{eq:dist_flow}
    \frac{\partial}{\partial \tau}{\boldsymbol{\phi}^{d}_{\rm b}}(\tau,\x) \! = \! \boldsymbol{f}_{\rm c l}(\phidb{\tau}{\x}) \!+ \!\boldsymbol{d}(\tau\! +\! t), \nspace{6} \boldsymbol{\phi}^{d}_{\rm b}(0,\x)\! = \!\x,
\end{align}
and the flow with the current disturbance estimate, $\boldsymbol{\phi}^{\hat{d}}_{\rm b}(\tau,\x)$:
\begin{align} \label{eq:est_dist_flow}
    \frac{\partial}{\partial \tau}{\boldsymbol{\phi}}^{\hat{d}}_{\rm b}(\tau,\x) = \boldsymbol{f}_{\rm cl}(\boldsymbol{\phi}^{\hat{d}}_{\rm b}(\tau,\x)) + \dhat(t), \nspace{8} \boldsymbol{\phi}^{\hat{d}}_{\rm b}(0,\x) = \x.
\end{align}
Notice that $\dhat(t)$ is a function of the global time $t$
rather than the backup time $\tau$, because the estimate of the disturbance cannot be updated over the flow (that would require future state data).
As such, this term is a constant over $\tau\in[0,T]$.

\begin{figure}[t]
    \centerline{\includegraphics[width=.95\columnwidth]{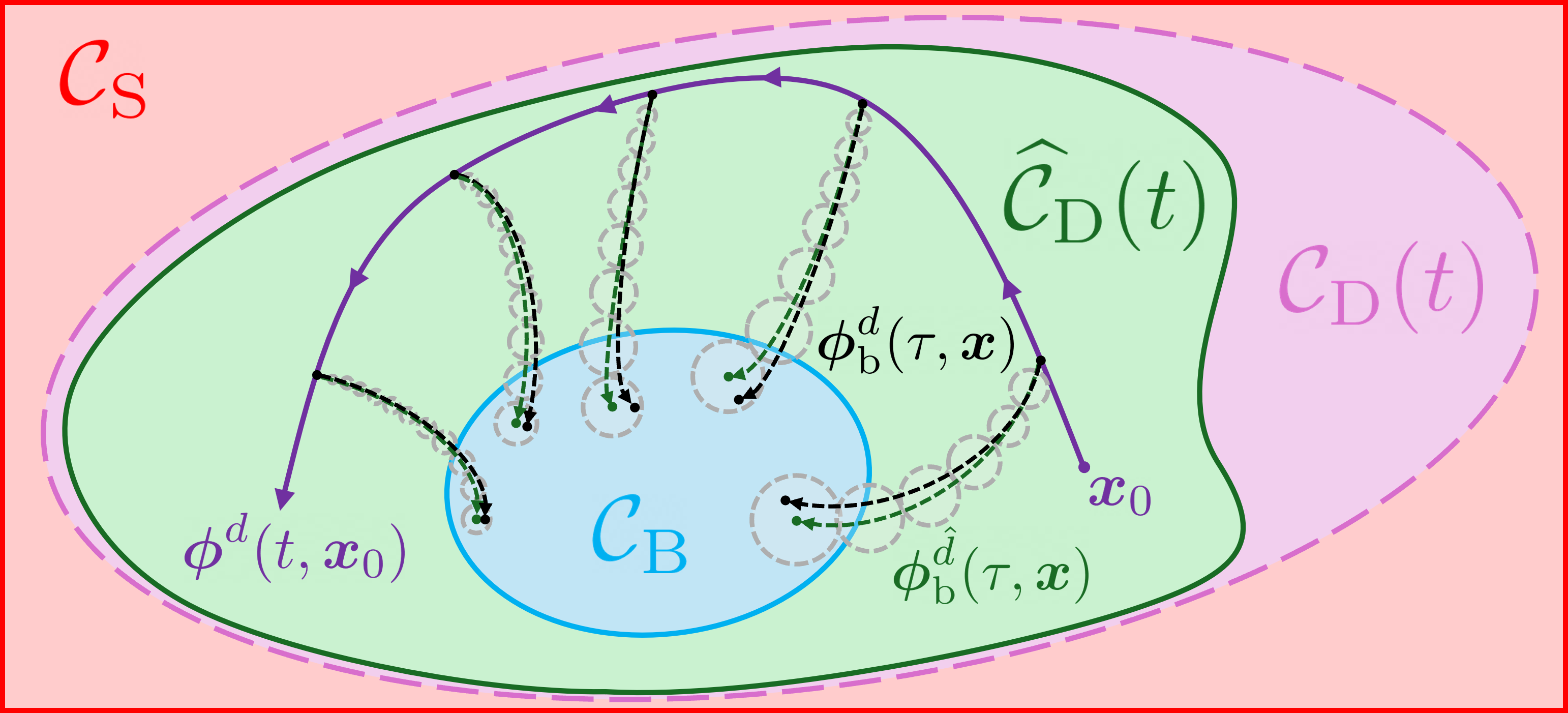}}
    \vspace{.2cm}
    \caption{Illustration of the presented robust safety-critical control method with an uncertainty estimator. The set $\Cdhat$, a known subset of an unknown controlled invariant set $\Cd$, is used to guarantee the safety of the disturbed flow $\boldsymbol{\phi}^{{d}}(t,\boldsymbol{x}_0)$. An uncertainty estimator
    shrinks the uncertainty bounds over time $t$.}
    \label{fig: overview}
\end{figure}

Consider next a time-varying set $\Cd \subseteq \Cs$ for all ${t \geq 0}$:
\begin{align} \label{eq: Cd}
    \hspace{-.09cm}
    \Cd \triangleq \left\{ \boldsymbol{x} \in \mathcal{X} \,\middle|\, 
    \begin{array}{c}
    h(\phid) \geq 0, \forall \nspace{1} \tau \in [0,T], \\
    h_{\rm b}(\phidT) \geq 0 \\
    \end{array}
    \right\}.
\end{align}
Using the definition of $\Cd$ and the corresponding robust backup controller $\ub$, we have the following result.
\begin{lemma}[\hspace{-0.01cm}{\cite[Lemma 3]{vanWijk_DRbCBF_24}}] \label{lemma: cd_inv}
    The set $\Cd$ is controlled invariant\footnote{A time-varying set $\mathcal{C}(t) \subset \mathbb{R}^n$ is controlled invariant if a controller $\boldsymbol{k}:\mathcal{X} \times \R \rightarrow \mathcal{U}$, $\boldsymbol{u}=\boldsymbol{k}(\boldsymbol{x},t)$ exists which renders $\mathcal{C}(t)$ forward invariant.} and
    the robust backup controller $\ub$ renders $\Cd$ forward invariant\footnote{$\nspace{-3}$\cite[Def. 4.10]{blanchini_set-theoretic_2015} A time-varying set ${\mathcal{C}(t) \subset \mathbb{R}^n}$ is forward invariant along \eqref{eq: disturbed_dyn} if for all $t_0$, ${\boldsymbol{x}(t_0) \in \mathcal{C}(t_0) \implies \boldsymbol{x}(t) \in \mathcal{C}(t)}$ for all ${t \geq t_0}$.} along \eqref{eq: disturbed_dyn} such that
    \begin{align}
        \nspace{-6}\boldsymbol{x}(0) \in \mathcal{C}_{\rm D}(0) \implies \boldsymbol{\phi}_{\rm b}^{d} (t, \boldsymbol{x}(0)) \in \mathcal{C}_{\rm D}(t), \forall \nspace{1} t \geq 0.
    \end{align}
\end{lemma}
These properties could allow one to feasibly enforce the forward invariance of $\Cd$.
However, the set $\Cd$ and the disturbed flow $\phid$ are unknown.
Instead, we use safety conditions for a known subset of $\Cd$, illustrated in Fig.~\ref{fig: overview}.

Consider a new time-varying set, $\Cdhat$:
\begin{align} \label{eq: Cdhat}
    \hspace{-.25cm}
    \Cdhat \triangleq \left\{ \boldsymbol{x} \in \mathcal{X} \,\middle|\, 
    \begin{array}{c}
    h(\phidhat) \geq \epsilon_\tau, \forall \nspace{1} \tau \in [0,T], \! \\
    h_{\rm b}(\phidhatT) \geq \epsilon_{\rm b} \\
    \end{array}
    \right\},
\end{align}
defined by the estimate flow and the tightening terms $\epsilon_{\rm \tau}$ and $\epsilon_{\rm b}$.
For carefully chosen tightening terms, $\Cdhat$ is a subset of $\Cd$. In the following Lemma, we provide an example of tightening terms which ensure that $\Cdhat \subseteq \Cd$.
\begin{lemma} \label{lemma: cdhat_subset}
    Let $\mathcal{L}_{h}$ and $\mathcal{L}_{h_{\rm b}}$ be the Lipschitz constants of $h$ and $h_{\rm b}$, respectively, and let ${\delta_{{\rm max}}(\tb,t)}$ be a norm bound on the deviation between $\phid$ and $\phidhat$ at backup time ${\tau \in [0,T]}$ and global time ${t \geq 0}$:
    \begin{equation}
        \norm{\phid - \phidhat} \leq \delta_{{\rm max}}(\tb,t),
        \label{eq:flow_bound}
    \end{equation}
    for all ${\boldsymbol{x} \in \Cs}$. If ${\epsilon_{\tb} \geq \mathcal{L}_{h} \delta_{{\rm max}}(\tb,t)}$ and ${\epsilon_{\rm b} \geq \mathcal{L}_{h_{\rm b}} \delta_{{\rm max}}(\Tt,t)}$ hold for all ${\tau \in [0,T]}$ and ${t \geq 0}$, then ${\Cdhat \!\subseteq\! \Cd}$.
    \begin{proof}
        Consider any state ${\boldsymbol{x} \in \Cdhat}$. Membership to $\Cdhat$ implies ${h(\phidhat) \geq \epsilon_{\tb} \geq \mathcal{L}_{h} \delta_{{\rm max}}(\tb,t)}$.
        It follows that
        \begin{align*}
             h(\phid) & = h(\phidhat) - \big( h(\phidhat) - h(\phid) \big) \\
             & \geq \mathcal{L}_{h} \delta_{{\rm max}}(\tb,t) - \big| h(\phidhat) - h(\phid) \big|.
        \end{align*}
        By using the Lipschitz continuity of $h$ and~\eqref{eq:flow_bound}, we have
        \begin{align*}
            | h (\phidhat) - h (\phid) |
            \leq \mathcal{L}_{h} \delta_{{\rm max}}(\tb,t),
        \end{align*}
        meaning that ${h(\phid) \geq 0}$ for any ${\boldsymbol{x} \in \Cdhat}$. Similarly, for any ${\boldsymbol{x} \in \Cdhat}$, the flow with the estimated disturbance satisfies ${h_{\rm b}(\phidhatT) \geq \epsilon_{\rm b}\geq \mathcal{L}_{h_{\rm b}} \delta_{{\rm max}}(\Tt,t)}$, and 
        \begin{align*}
            | h_{\rm b} (\phidhatT) - h_{\rm b}(\phidT) | \leq \mathcal{L}_{h_{\rm b}} \delta_{{\rm max}}(\Tt,t).
        \end{align*}
        This guarantees that ${h_{\rm b}(\phidT) \geq 0}$.
        Thus, based on~\eqref{eq: Cd},
        ${\boldsymbol{x} \in \Cd}$ holds for any ${\boldsymbol{x} \in \Cdhat}$, so ${\Cdhat \subseteq \Cd}$. 
    \end{proof}
\end{lemma}

Similar to as in \cite[Lemma 1]{vanWijk_DRbCBF_24}, Lemma~\ref{lemma: cdhat_subset} guides the selection of $\epsilon_{\rm \tau}$ and $\epsilon_{\rm b}$ using a bound $\delta_{\rm max}(\tau,t)$ on the discrepancy between the unknown (disturbed) and estimated flows, and the Lipschitz constants of the barriers $h$ and $h_{\rm b}$. We note however, that in cases where $h$ and $h_{\rm b}$ have a specific structure (e.g., are linear, quadratic, and/or convex), tighter $\epsilon_{\tau}$ and $\epsilon_{\rm b}$ terms may be obtained as shown in \cite[Remark 2]{vanwijk2026outputfeedbackbackupcontrol}.
However, all such forms of tightening terms rely on a characterization of the uncertainty in the flow prediction, captured by \eqref{eq:flow_bound}. For this work, obtaining such a bound on the flow deviation requires characterizing the fidelity of the disturbance estimate.
\begin{lemma} \label{lemma: dob_backup_bound}
    Given a disturbance satisfying Assumption~\ref{as:disturbance} and an error-quantified uncertainty estimator as in Definition~\ref{def:EQUE} with error bound $\Bar{e}(t)$, we have
    \begin{align}
        \|{\boldsymbol{d}(\tau + t) - \dhat(t)}\| \leq \delta_v \tau + \Bar{e}(t),
        \label{eq:disturbance_bound}
    \end{align}
    for any global time $t \geq 0$ and backup time $\tau \geq 0$.
    \begin{proof}
        The triangle inequality implies for any ${t,\tau \geq 0}$ that
        \begin{equation*} \label{eq: dist_fixed}
        \begin{aligned} 
        \|\boldsymbol{d}(\tau + t) - \hat{\boldsymbol{d}}(t)\| 
        &\leq \|\boldsymbol{d}(\tau + t) - \boldsymbol{d}(t)\| + \|\boldsymbol{d}(t) - \hat{\boldsymbol{d}}(t)\|, 
        \end{aligned}
        \end{equation*}
        where
        ${\|\boldsymbol{d}(t) - \hat{\boldsymbol{d}}(t)\| \leq \bar{e}(t)}$
        according to Definition~\ref{def:EQUE},
        and
        \begin{align*}
        \|\boldsymbol{d}(\tau + t) - \boldsymbol{d}(t)\| \leq \int_{t}^{\tau + t} \|\dot{\boldsymbol{d}}(s)\| \, ds \leq \delta_v\tau,
        \end{align*}
        because $\|\dot{\boldsymbol{d}}(t)\| \leq \delta_v$.
        These inequalities lead to~\eqref{eq:disturbance_bound}.
    \end{proof}
\end{lemma}
Using this result, we provide two examples of general flow bounds which could be used to satisfy \eqref{eq:flow_bound} in Lemma~\ref{lemma: cdhat_subset}.
\begin{lemma} \label{lemma: delta_max}
    For systems \eqref{eq:dist_flow} and \eqref{eq:est_dist_flow} let $\boldsymbol{f}_{\rm cl}$ be locally Lipschitz on $\mathcal{X}$ with Lipschitz constant $\mathcal{L}_{\rm cl}$ where $\mathcal{X}$ is compact.
    If Assumption~\ref{as:disturbance} is satisfied, then~\eqref{eq:flow_bound} holds for all $\tau \in [0,T]$ and $t \geq 0$ with
    \begin{align} \label{eq: delta_max_bound}
    \delta_{\max}(\tau,t) \triangleq \bigg(\frac{\delta_v}{\mathcal{L}_{\rm cl}^2} + \frac{\bar{e}(t)} {\mathcal{L}_{\rm cl}} \bigg) \Big( {\rm e}^{\mathcal{L}_{\rm cl}\tau} - 1 \Big) - \frac{\delta_v}{\mathcal{L}_{\rm cl}} \tau.
    \end{align}
     \begin{proof}
        Introducing ${\Delta \phi(\tau,\boldsymbol{x}) \triangleq \|\phid - \phidhat\|}$
        and expanding the bound between the flows~\eqref{eq:dist_flow} and~\eqref{eq:est_dist_flow},
        \begin{align*}
            \Delta \phi(\tau,\boldsymbol{x})
            \leq \!\int^{\tau}_{0}\! \mathcal{L}_{\rm cl}\, \Delta \phi(s,\boldsymbol{x}) ds
            + \!\int^{\tau}_{0}\! \big\|\boldsymbol{d}(s + t) - \hat{\boldsymbol{d}}(t) \big\| ds.
        \end{align*}
        Applying Lemma~\ref{lemma: dob_backup_bound}, we obtain
        \begin{align*}
            \Delta \phi(\tau,\boldsymbol{x})
            \leq \int^{\tau}_{0} \mathcal{L}_{\rm cl}\, \Delta \phi(s,\boldsymbol{x}) ds
            + \frac{\delta_v}{2} \tau^2 + \bar{e}(t) \tau.
        \end{align*}
        Using the Gr\"onwall--Bellman Inequality \cite[Lemma 2.1]{khalil2002nonlinear},
        \begin{align*}
            \Delta \phi(\tau,\boldsymbol{x}) \!\leq\! {\!\int^{\tau}_{0}\!\! \!\Big( \frac{\delta_v}{2} s^2 \!+\! \bar{e}(t) s \Big) \mathcal{L}_{\rm cl} {\rm e}^{\mathcal{L}_{\rm cl} (\tau - s)} ds \!+\! \frac{\delta_v}{2} \tau^2 \!+\! \bar{e}(t) \tau}.
        \end{align*}
        Calculating the integral yields the result.
    \end{proof}
\end{lemma}

Alternatively, it is often possible to obtain an upper bound on the logarithmic norm (or \textit{matrix measure}) of closed-loop nonlinear systems \cite{Sontag2010}. This bound is typically tighter than using Lipschitz arguments on the closed-loop system.

\begin{lemma} \label{lemma: logBound}
Let the induced logarithmic norm\footnote{Let the {induced logarithmic norm} of a matrix ${\boldsymbol{A} \in \mathbb{R}^{n \times n}}$ be denoted ${\mu(\boldsymbol{A})}$, defined by ${\mu(\boldsymbol{A}) = \lim_{z \rightarrow 0^+} \frac{\norm{\boldsymbol{I} + z \boldsymbol{A}} - 1}{z}}.$} of the closed-loop Jacobian ${\jac}$ defined in \eqref{eq:F_cl_def} be bounded by a constant ${c \in \mathbb{R}}$ s.t. ${\mu(\jac(\x)) \leq c}$ for all ${\x \in \mathcal{X}}$.
If Assumption~\ref{as:disturbance} is satisfied, then~\eqref{eq:flow_bound} holds for all ${\tau \in [0,T]}$ and ${t \geq 0}$ with 
\begin{align} \label{eq:lognorm_bound}
    \delta_{\rm max}(\tau,t) \triangleq \bigg(\frac{\delta_v}{c^2} + \frac{\bar{e}(t)}{c} \bigg) \Big( {\rm e}^{c\tau} - 1 \Big) - \frac{\delta_v}{c} \tau.
\end{align}
\begin{proof}
    Because $\dhat(t)$ and ${\boldsymbol{d}(\tau + t)}$ have no explicit state dependence, the closed-loop Jacobian of ${{\frac{\partial}{\partial {\tau}}}{\boldsymbol{\phi}^{d}_{\rm b}}(\tau,\x)}$ in \eqref{eq:dist_flow} and ${{\frac{\partial}{\partial {\tau}}}{\boldsymbol{\phi}}^{\hat{d}}_{\rm b}(\tau,\x)}$ in \eqref{eq:est_dist_flow} is the same as the Jacobian for the \textit{unperturbed} backup flow \eqref{eq: nomFlow}. Thus, applying \cite[Corollary 3.17]{contraction_bullo} under the assumption that ${\mu(\jac(\x)) \leq c}$ for all $\x\in\mathcal{X}$, and using Lemma~\ref{lemma: dob_backup_bound} we have
    \begin{align*}
        \Delta \phi(\tau,\boldsymbol{x}) \leq \underbrace{\Delta \phi(0,\boldsymbol{x})}_{=0} {\rm e}^{c\tau} + \int^{\tau}_0 {\rm e}^{c(\tau - s)} \big( \delta_v s + \Bar{e}(t) \big) \nspace{2} ds.
    \end{align*}
    Integrating by parts from $0$ to $\tau$ yields the result.
\end{proof}
\end{lemma}
\begin{remark}
    While the flow bounds $\dmax(\tau,t)$ in Lemma~\ref{lemma: delta_max} and Lemma~\ref{lemma: logBound} grow with $\tau$, they shrink with $t$ if
    the estimate of the disturbance improves as $t$ increases (i.e., $\bar{e}(t)$
    decreases).
    The bounds in \eqref{eq: delta_max_bound} and \eqref{eq:lognorm_bound} are general, and there may exist tighter problem-specific bounds (e.g., for linear systems, or where differential flatness may be leveraged).
    In the case when ${c < 0}$ in \eqref{eq:lognorm_bound}, the closed-loop backup dynamics are said to be contracting \cite{contraction_lcss_bullo24,contraction_bullo,LOHMILLER1998683} which yields tight flow bounds \cite{vanWijk_DRbCBF_24}.
\end{remark}
We can now state our main result about the set $\Cdhat$, which is comprised only of known terms. 
\begin{theorem} \label{thm: controlInv_cd}
Let $\epsilon_{\tb}$ and $\epsilon_{\rm b}$ satisfy ${\epsilon_{\tb} \geq \mathcal{L}_{h} \delta_{{\rm max}}(\tb,t)}$ and ${\epsilon_{\rm b} \geq \mathcal{L}_{h_{\rm b}} \delta_{{\rm max}}(\Tt,t)}$ for all ${\tau \in [0,T]}$ and ${t \geq 0}$, with ${\delta_{{\rm max}}(\tb,t)}$ satisfying \eqref{eq:flow_bound} (such as those defined in~\eqref{eq: delta_max_bound} or \eqref{eq:lognorm_bound}).
For any ${\boldsymbol{x} \in \Cdhat}$, there exists a controller $\boldsymbol{k}(\x)$ such that ${\boldsymbol{\phi}^{d} (\vartheta, \boldsymbol{x}) \in \mathcal{C}_{\rm D}(t + \vartheta)\subseteq \Cs, \forall \nspace{1} \vartheta \geq 0}$. 
\begin{proof}
By Lemma~\ref{lemma: cdhat_subset}, ${\boldsymbol{x} \in \Cdhat \nspace{-6} \implies \boldsymbol{x} \in \Cd}$, and by Lemma~\ref{lemma: cd_inv}, $\boldsymbol{k}_{\rm b}$ ensures ${\boldsymbol{\phi}^{d} (\vartheta, \boldsymbol{x}) \in \mathcal{C}_{\rm D}(t + \vartheta)}$, ${\forall \nspace{2} \vartheta \geq 0}$.
\end{proof}
\end{theorem}

\subsection{Robust Safety-Critical Control}

We are now ready to derive control conditions to ensure the robust safety of \eqref{eq: disturbed_dyn} via the forward invariance of $\Cdhat$ (where ${\Cdhat \subseteq \Cd \subseteq \Cs}$). From~\eqref{eq: Cdhat}, this requires
\begin{align} \label{eq:invcond_1}
    \begin{gathered}
    \dot{h}(\phidhat, \boldsymbol{u}) - \frac{\partial \epsilon_\tau}{\partial t} \ge - \alpha \big( h(\phidhat) - \epsilon_\tau \big), \\
    \dot{h}_{\rm b}(\phidhatT, \boldsymbol{u}) - \frac{\partial \epsilon_{\rm b}}{\partial t} \ge -\alpha_{\rm b} \big( h_{\rm b}(\phidhatT) - \epsilon_{\rm b} \big),
    \end{gathered}
\end{align}
where $\epsilon_{\tau}$ and $\epsilon_{\rm b}$ account for the discrepancy between the true (unknown) backup flow and its estimate.
Note that $\epsilon_{\tau}$ and $\epsilon_{\rm b}$ are functions of both $t$ and $\tau$. 

\begin{lemma} \label{lemma:total_derivs}
    The terms ${\dot{h}(\phidhat, \boldsymbol{u})}$ and ${\dot{h}_{\rm b}(\phidhatT, \boldsymbol{u})}$ are given by
    \begin{subequations}
        \begin{align*}
            \dot{h}(\phidhat, \boldsymbol{u}) \!&=\! \nabla h(\phidhat) \big( \stmdhat \dot{\boldsymbol{x}} \!+\! \boldsymbol{\Theta}(\tau,\x) \skew{7}{\dot}{\skew{7}{\hat}{\bm{d}}} \nspace{1}\big), \\
            \dot{h}_{\rm b}(\phidhatT, \boldsymbol{u}) \!&=\! \nabla h_{\rm b}(\phidhatT) \big( \stmdhatT \dot{\boldsymbol{x}} \!+\! \boldsymbol{\Theta}(T,\x) \skew{7}{\dot}{\skew{7}{\hat}{\bm{d}}} \nspace{1} \big),
        \end{align*}
    \end{subequations}
    with ${{{\boldsymbol{\Phi}}}(\tau, \boldsymbol{x}) \!\triangleq\! {\partial \phidhat}/{\partial \x}}$ and ${\boldsymbol{\Theta}(\tau,\x) \!\triangleq\! {\partial \phidhat}/{\partial \dhat}}$ 
    which evolve with
    \begin{subequations} \label{eq:dobcbf_stm_all}
    \begin{flalign} \label{eq:dobcbf_stmProp}
    &\frac{\partial}{{\partial} \tau} {{\boldsymbol{\Phi}}}(\tau, \boldsymbol{x}) \! = \! \jac(\phidhat) {{\boldsymbol{\Phi}}}(\tau, \boldsymbol{x}), \nspace{30}
    \boldsymbol{\Phi}(0,\boldsymbol{x}) = {\boldsymbol{I}}, && \raisetag{\baselineskip}\\
    \label{eq: Theta_stm}
    &\frac{\partial}{{\partial} \tau}{{\boldsymbol{\Theta}}}(\tau, \boldsymbol{x}) \! = \! \jac(\phidhat)\boldsymbol{\Theta}(\tau,\x) \!+ \!\eye,
    \nspace{6}
    \boldsymbol{\Theta}(0,\boldsymbol{x}) \! = \! {\boldsymbol{0}}. && \raisetag{\baselineskip}
    \end{flalign}        
    \end{subequations}
    \begin{proof}
        See Appendix \ref{proof:total_derivs}.
    \end{proof}
\end{lemma}

\begin{remark}
In addition to forward integrating the sensitivity $\boldsymbol{\Phi}$ of the flow with respect to changes in the initial condition $\x$ using~\eqref{eq:dobcbf_stmProp} (similar to~\eqref{eq: stm_nominal} in the standard bCBF approach), we also forward integrate the sensitivity $\boldsymbol{\Theta}$ of the flow with respect to changes in the disturbance estimate $\dhat$ using \eqref{eq: Theta_stm}.
The sensitivity term $\boldsymbol{\Theta}$ is crucial to achieve robust safety against uncertainty estimation errors and, importantly, its formulation in~\eqref{eq: Theta_stm} is applicable generally for arbitrary error-quantified uncertainty estimators. Therefore, the proposed approach offers significantly more utility than merely combining two existing approaches such as the disturbance observer in~\cite{das2025robustcontrolbarrierfunctions} and the robust safety-critical controller in~\cite{vanWijk_DRbCBF_24}.
\end{remark}
\begin{remark}
In~\eqref{eq: Theta_stm}, we essentially consider $\dhat$ as an auxiliary state in order to compute the flow gradient. 
This method of computing the flow gradient is advantageous because ${{{\boldsymbol{\Phi}}}(\tau, \boldsymbol{x})}$ and ${\boldsymbol{\Theta}(\tau,\x)}$ can be computed \textit{exactly}\footnote{Alternatively, the term ${(\partial \phidhat/\partial \dhat) \dhatdot}$ may be captured by ${\partial \phidhat/\partial t}$, and this term can be approximated directly using numerical techniques such as finite differencing. While this would only require integrating ${n + n^2}$ ODEs (like standard bCBFs in Section~\ref{sec:bCBF}), the safety guarantees may degrade with the precision of the approximation of ${\partial \phidhat/\partial t}$.}. However, this requires integrating ${n + 2n^2}$ ordinary differential equations (ODEs) given by~\eqref{eq:est_dist_flow} and~\eqref{eq:dobcbf_stm_all} rather than the ${n + n^2}$ ODEs in \eqref{eq: nomFlow} and~\eqref{eq: stm_nominal} required by the standard bCBF method\footnote{To reduce the number of ODEs required, one may also use techniques to compute only certain components of the gradient, as discussed in \cite{vanwijk2025safetycriticalcontrolboundedinputs}.}.  
\end{remark}

With the sensitivity terms established, we can now substitute them into the safety constraints.
Applying Lemma~\ref{lemma:total_derivs} to \eqref{eq:invcond_1} 
yields
\begin{align} \label{eq: nagumo_cdhat}
    \begin{gathered}
    \hspace{-1.1cm}\nabla h(\phidhat) \big( \stmdhat \dot{\boldsymbol{x}} + \boldsymbol{\Theta}(\tau,\x) \skew{7}{\dot}{\skew{7}{\hat}{\bm{d}}} \nspace{1}\big) \ge \\
    \hspace{2cm}
    - \alpha \big( h(\phidhat) - \epsilon_\tau \big) + \frac{\partial \epsilon_\tau}{\partial t}, \\ 
    \hspace{-.5cm} 
    \nabla h_{\rm b}(\phidhatT) \big( \stmdhatT \dot{\boldsymbol{x}} + \boldsymbol{\Theta}(T,\x) \skew{7}{\dot}{\skew{7}{\hat}{\bm{d}}} \nspace{1} \big) \ge \\ 
    \hspace{2.5cm}
    -\alpha_{\rm b} \big( h_{\rm b}(\phidhatT) - \epsilon_{\rm b} \big) + \frac{\partial \epsilon_{\rm b}}{\partial t},
    \end{gathered}
\end{align}
for all ${\tau \in [0,T]}$, where ${\dot{\boldsymbol{x}} \!=\! \boldsymbol{f}(\x) \!+\! \boldsymbol{g}(\x)\boldsymbol{u} \!+\! \boldsymbol{d}}$.
The derivative
$\skew{7}{\dot}{\skew{7}{\hat}{\bm{d}}}$ is obtained from~\eqref{eq:uncertainty_estimator}.
For example, for the uncertainty estimator in \eqref{eq:DOB_all} we have ${\skew{7}{\dot}{\skew{7}{\hat}{\bm{d}}} \!=\! \bLam \big(\dot{\x} \!-\! \dot{\bm{\xi}} \big) \!=\! \bLam \big(\boldsymbol{d} \!-\! \dhat \big)}$. 

Enforcing constraint~\eqref{eq: nagumo_cdhat} could ensure the forward invariance of $\Cdhat$ and thereby guarantee safety.
However,~\eqref{eq: nagumo_cdhat} includes the unknown disturbance $\boldsymbol{d}$ in $\dot{\boldsymbol{x}}$ and $\dhatdot$.
Thus, we derive sufficient conditions for the satisfaction of~\eqref{eq: nagumo_cdhat} with a method inspired by \cite{jankovic_robust_2018}.
The following Theorem establishes that controllers satisfying these conditions ensure the robust safety of the system~\eqref{eq: disturbed_dyn} despite the unknown disturbance.

\begin{theorem} \label{thm: mainResult}
    Any locally Lipschitz controller ${\boldsymbol{k}:\mathcal{X} \rightarrow \mathcal{U}}$, ${\boldsymbol{u}=\boldsymbol{k}(\boldsymbol{x})}$ satisfying
    \begin{subequations} \label{eq: mainthmConstraints}
        \begin{align} 
            \! &\begin{aligned}
            \nabla h(\phidhat) \stmdhat \big( \boldsymbol{f}(\boldsymbol{x}) + \boldsymbol{g}(\boldsymbol{x}) \boldsymbol{u} + \dhat \nspace{1}\big) \ge \\ 
             - \alpha \big( h(\phidhat) - \epsilon_{\tau} \big) + \frac{\partial \epsilon_\tau}{\partial t} + \rho,  \label{eq: mainthmConstraints_cont}
             \end{aligned} \\ 
            \! & \begin{aligned}
            \nabla h_{\rm b}(\phidhatT) \stmdhatT \big( \boldsymbol{f}(\boldsymbol{x}) + \boldsymbol{g}(\boldsymbol{x}) \boldsymbol{u}+ \dhat \nspace{1}\big) \ge \\
             - \alpha_{\rm b} \big( h_{\rm b}(\phidhatT) - \epsilon_{\rm b} \big) + \frac{\partial \epsilon_{\rm b}}{\partial t} + \rho_{\rm b},  \label{eq: mainthmConstraints_reach}
             \end{aligned} 
        \end{align}
    \end{subequations}
    for all ${\tau \in [0,T]}$, ${t \geq 0}$, $\boldsymbol{x} \in \Cs$, and  robustness terms
    \begin{align}
    \begin{gathered}
        \rho \geq -\nabla h(\phidhat) \big( \stmdhat \boldsymbol{e} + \boldsymbol{\Theta}(\tau,\x)\dhatdot \nspace{2} \big), \\
        \rho_{\rm b} \geq -\nabla h_{\rm b}(\phidhatT) \big( \stmdhatT \boldsymbol{e} + \boldsymbol{\Theta}(T,\x)\dhatdot \nspace{2}\big), \label{eq:robust_general} 
    \end{gathered}
    \end{align}
    renders the set ${\Cdhatcon(t) \!\subseteq\! \Cdcon(t) \!\subseteq\! \Cs}$ forward invariant for~\eqref{eq: disturbed_dyn}.
    \begin{proof}
         Substituting ${\boldsymbol{d}=\dhat + \boldsymbol{e}}$ into \eqref{eq: nagumo_cdhat}, the conditions in \eqref{eq: mainthmConstraints} imply that \eqref{eq: nagumo_cdhat} holds because the robustness terms overapproximate the terms associated with the unknown values ${\boldsymbol{d}}$ and ${\dhatdot}$. Applying Theorem~\ref{thm: cbf} to system \eqref{eq: disturbed_dyn}, the satisfaction of \eqref{eq: nagumo_cdhat} yields
         the forward invariance of $\Cdhatcon(t)$.
         By Lemma~\ref{lemma: cdhat_subset}, we have ${\Cdhatcon(t) \subseteq \Cdcon(t)}$.
    \end{proof}
\end{theorem}

\begin{corollary}
    For the uncertainty estimator in~\eqref{eq:DOB_all}, condition~\eqref{eq:robust_general} in Theorem~\ref{thm: mainResult} can be satisfied with robustness terms
    \begin{align} 
    \begin{gathered}
        \rho = \bar{e}(t) \norm{\nabla h(\phidhat) \big( \stmdhat + \boldsymbol{\Theta}(\tau,\x)\bLam \big)}, \\
        \rho_{\rm b} = \bar{e}(t) \norm{\nabla h_{\rm b}(\phidhatT) \big( \stmdhatT + \boldsymbol{\Theta}(T,\x)\bLam \big)}. \label{eq:robust_DOB}
    \end{gathered}
    \end{align}
    \begin{proof}
        First, substitute ${\dhatdot=\bLam \boldsymbol{e}}$ into \eqref{eq:robust_general} and collect like terms with $\boldsymbol{e}$. Then, using the Cauchy--Schwarz inequality and the upper bound ${\|\boldsymbol{e}(t)\| \leq \bar{e}(t)}$ from Lemma~\ref{lemma: dob_errorBound}, it is clear that \eqref{eq:robust_DOB} implies the satisfaction of \eqref{eq:robust_general}.
    \end{proof}
\end{corollary}

Theorem~\ref{thm: mainResult} can now be used to develop a novel point-wise optimal safe controller via the proposed \textit{Uncertainty Estimator Backup CBF (UE-bCBF)} approach:
\begin{align*} 
    \boldsymbol{k}^\star
    (\boldsymbol{x},t) = \underset{\boldsymbol{u} \in \mathcal{U}}{\text{arg\,min}} \mkern9mu &
    \left\Vert \boldsymbol{k}_{\rm p}(\x)-\boldsymbol{u}\right\Vert^{2} \quad \tag{UE-bCBF-QP} \label{eq:dob-bcbf-qp} \\
    \text{s.t.  } 
    & \eqref{eq: mainthmConstraints_cont}, \ \eqref{eq: mainthmConstraints_reach},
\end{align*}
for all ${\tau \in [0, \Tt]}$ and ${t \geq 0}$. Because $\Cdhat$ is itself not controlled invariant, the \eqref{eq:dob-bcbf-qp} may technically experience infeasibility issues. However, from Lemma~\ref{lemma: cd_inv}, $\Cd$ is controlled invariant, and thus the robust backup controller $\boldsymbol{k}_{\rm b}$ satisfying Assumption~\ref{assump: robust_inv} can be used to stay in $\Cd$, guaranteeing robust safety under input bounds since ${\Cdhat \subseteq \Cd \subseteq \Cs}$ and ${\ub(\boldsymbol{x}) \in \mathcal{U}}$ for all ${\x \in \Cd}$, ${t \geq 0}$. Such a controller could take the form
\begin{align*} 
    \boldsymbol{k}_{\rm safe}(\boldsymbol{x},t) = \mathds{1}_{\mathcal{F}}( \boldsymbol{x},t)\boldsymbol{k}^\star(\boldsymbol{x},t) + \big(1 \!-\! \mathds{1}_{\mathcal{F}}(\boldsymbol{x},t)\big)\boldsymbol{k}_{\rm b}(\x).
\end{align*}
Here, $\mathds{1}_{\mathcal{F}}$ is the indicator function defined such that ${\mathds{1}_{\mathcal{F}}(\boldsymbol{x},t) = {1}}$ if ${\exists \, \boldsymbol{u} \in \mathcal{U}}$ such that \eqref{eq: mainthmConstraints_cont} and \eqref{eq: mainthmConstraints_reach} hold, and $0$ otherwise.
Alternatively, a smooth switching approach may be used as in \cite{rabiee_softmin_bcbf}.

%% file: sections/05_NumericalExamples.tex
\section{Numerical Examples}\label{sec:examples}

\begin{figure}
\centering
\begin{subfigure}{.9\columnwidth}
\centerline{\includegraphics[width=.95\columnwidth]{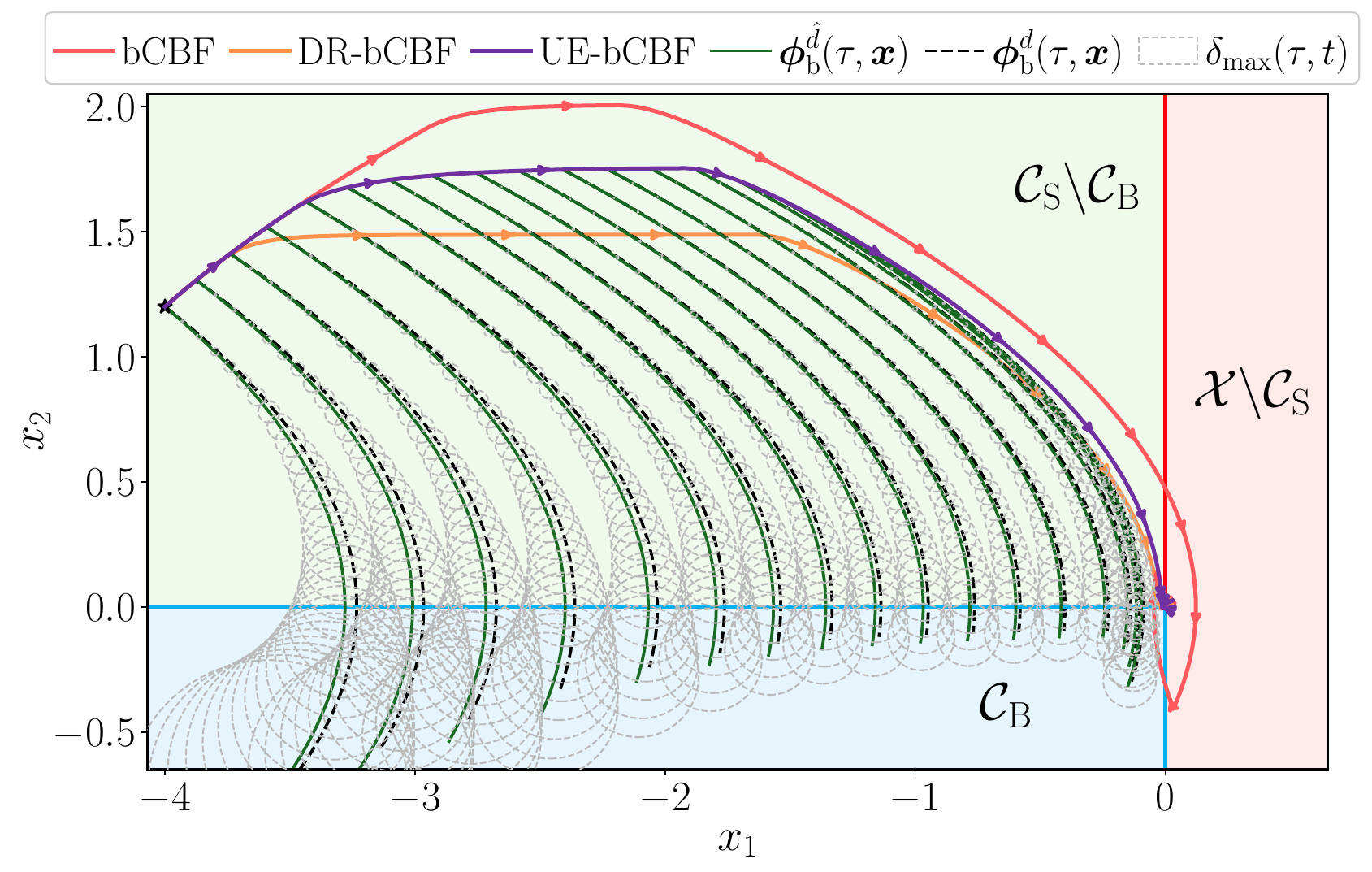}}
\end{subfigure}
\begin{subfigure}{.95\columnwidth}
\centerline{\includegraphics[width=.95\columnwidth]{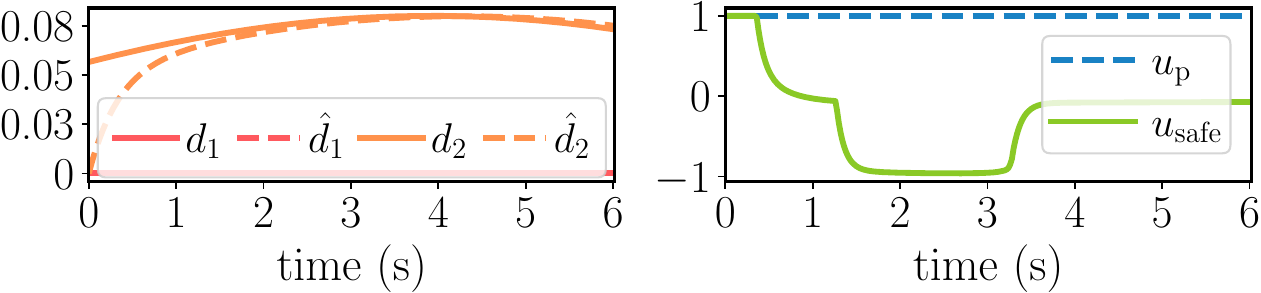}}
\vspace{.2cm}
\end{subfigure}
\caption{Simulation of the double integrator~\eqref{eq: db_int} with ${\omega=0.2}$ using the proposed uncertainty estimator backup CBF controller \eqref{eq:dob-bcbf-qp}. The trajectory of the system (purple) indicates safe behavior despite the unknown disturbance (\textbf{top}). The controller uses estimated backup trajectories (green) that approximate the unknown backup flow under the true disturbance (black dashed). The flow uncertainty decreases over time $t$ thanks to the uncertainty estimator (see the gray circles centered on the estimated trajectories representing the Gr\"onwall norm balls from Lemma~\ref{lemma: delta_max}).
Indeed, the true disturbance is captured by its estimate (\textbf{bottom left}), while the control input stays bounded (\textbf{bottom right}).}
\label{fig:db_int_0.2}
\end{figure}

In this section, we demonstrate the effectiveness of the proposed approach using two simulation examples\footnote{Code available at: \href{https://github.com/davidvwijk/UE-bCBF}{https://github.com/davidvwijk/UE-bCBF}}.

\begin{casestudy} \label{ex:1}
Consider a double integrator given by
\begin{align} \label{eq: db_int}
    \dot{\boldsymbol{x}} = 
    [x_2 ~ u]^\top
    \nspace{-2}
    + \boldsymbol{d}(t),
\end{align}
with position $x_1$, velocity $x_2$, state ${\boldsymbol{x} \!=\! 
[x_1 ~ x_2]^\top \!\in\! \mathcal{X}\subset \mathbb{R}^2}$, and control input ${u \in \mathcal{U} = [-1, 1]}$. The safe set is defined as 
${\Cs \!\triangleq\! \{\boldsymbol{x} \!\in\! \mathcal{X} \!:\! -x_1 \!\geq\! 0 \}}$.
The unknown disturbance is time-varying, with
${\boldsymbol{d}(t) \!=\! \delta_d\nspace{1}[0~{\rm sin}{(\nspace{-1}\omega t \!+\! \frac{\pi}{4})}]^\top}$,
and we use the known bounds ${\norm{\boldsymbol{d}} \!\leq\! \delta_d \!=\! 0.08}$ and ${\|{\dot{\boldsymbol{d}}\|} \leq \delta_v \!=\! \delta_d \omega}$ for control design. The backup control law ${\boldsymbol{k}_{\rm b}(\boldsymbol{x}) \!=\! -1}$ brings the system to the backup set ${\Cb \!\triangleq\! \{\boldsymbol{x} \!\in\! \mathcal{X} \!:\! -x_1 \!\geq\! 0, -x_2 \!\geq\! 0 \}}$.
The primary controller, ${\boldsymbol{k}_{\rm p}(\boldsymbol{x}) \!=\! 1}$, is intentionally designed to drive \eqref{eq: db_int} to the unsafe right half-plane.

We simulate~\eqref{eq: db_int} with the proposed~\eqref{eq:dob-bcbf-qp} controller and the uncertainty estimator~\eqref{eq:DOB_all}, and we compare our approach with two baselines: the disturbance-robust backup CBF (DR-bCBF) solution in \cite{vanWijk_DRbCBF_24}, that is designed for the worst-case disturbance without utilizing an uncertainty estimator, and the standard~\eqref{eq:bcbf-qp} reviewed in Section~\ref{sec:bCBF}, that ignores the disturbance all together.
The results are shown in Fig.~\ref{fig:db_int_0.2} for ${\omega=0.2}$ and in Fig.~\ref{fig:db_int_0.0} for ${\omega=0}$. Both configurations indicate that the proposed UE-bCBF approach guarantees safety despite the unknown disturbance, and is less conservative (allowing higher velocity $x_2$) than the DR-bCBF. In contrast, the bCBF violates safety due to the disturbance.
We also depict the disturbed flow $\phid$, the estimated flow $\phidhat$, and its uncertainty bound from Lemma~\ref{lemma: delta_max} represented as circles.
As time $t$ goes on, the disturbance estimate gets more accurate and the circles shrink.
For ${\omega = 0}$, the uncertainty vanishes completely, implying that the set $\Cdhat$ approaches $\Cd$, since the disturbance is constant and the estimation error converges to zero by Lemma~\ref{lemma: dob_errorBound}. 

\begin{figure}
\centering
\begin{subfigure}{.9\columnwidth}
\centerline{\includegraphics[width=.95\columnwidth]{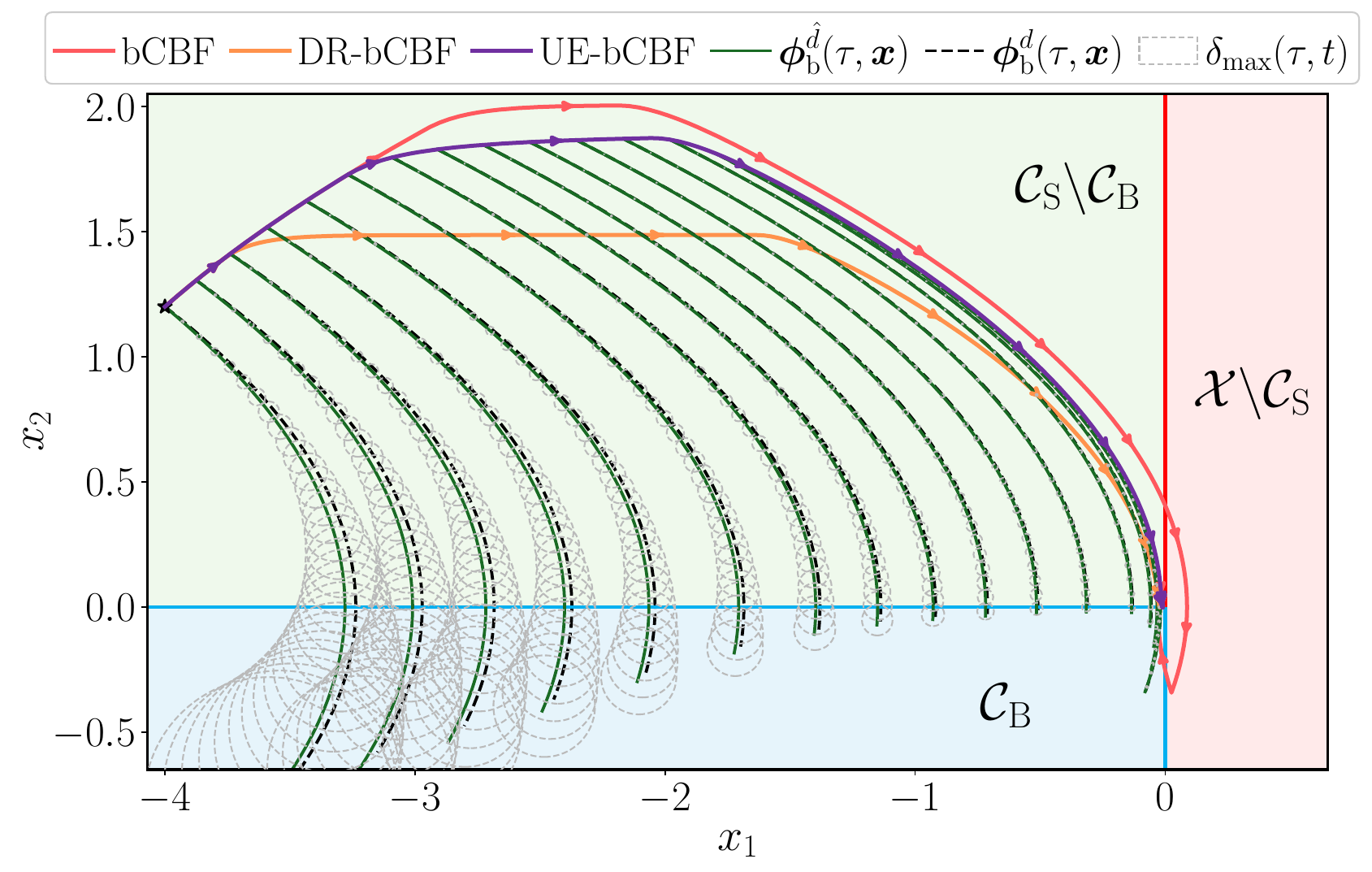}}
\end{subfigure}
\begin{subfigure}{.95\columnwidth}
\centerline{\includegraphics[width=.95\columnwidth]{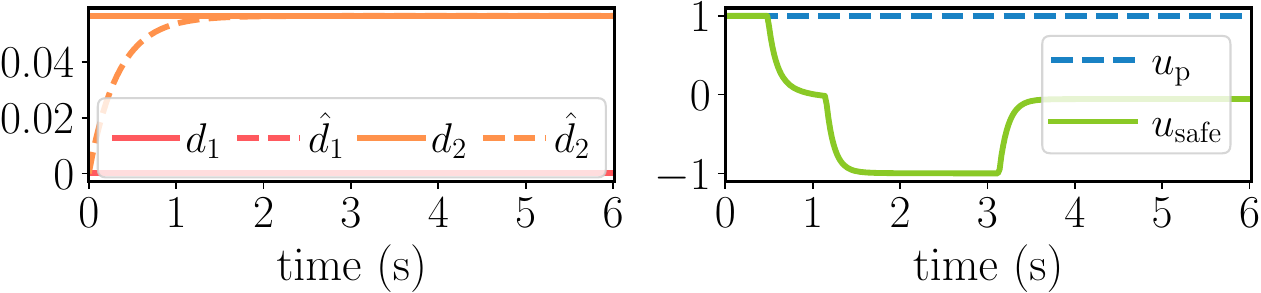}}
\end{subfigure}
\vspace{.2cm}
\caption{Simulation of \eqref{eq: db_int} with ${\omega=0}$ using the uncertainty estimator backup CBF controller \eqref{eq:dob-bcbf-qp}. 
}
\label{fig:db_int_0.0}
\end{figure}

\end{casestudy}

\begin{casestudy} \label{ex:2}
Consider next a planar quadrotor\cite{wu_acc_quadrotor2016}:
\begin{align}
    \!\underbrace{\begin{bmatrix}
      \dot{x}\\
      \dot{z}\\
      \dot{\theta}\\
      \ddot{x} \\
      \ddot{z} \\
      \ddot{\theta}
    \end{bmatrix}}_{\dot{\x}}
    \!=\! 
    \underbrace{\begin{bmatrix}
      \dot{x}\\
      \dot{z}\\
      \dot{\theta}\\
      0 \\
      -g_{\rm D} \\
      0
    \end{bmatrix}}_{\boldsymbol{f}(\boldsymbol{x})}
    \!+\!
    \underbrace{\begin{bmatrix}
        0 & 0 \\ 
        0 & 0 \\
        0 & 0 \\
        {\rm sin}(\theta)/m & 0 \\
        {\rm cos}(\theta)/m & 0 \\
        0 & -1/J \\
    \end{bmatrix}}_{\boldsymbol{g}(\x)}\!\!
    \underbrace{\begin{bmatrix}
        F \\
        M
    \end{bmatrix}}_{\boldsymbol{u}}
    \!+\! 
    \underbrace{\begin{bmatrix}
        0 \\
        0 \\
        0 \\
        d_4(t) \\
        d_5(t) \\
        0
    \end{bmatrix}}_{\boldsymbol{d}(t)}\!,
    \label{eq:quadrotor}
\end{align}
where $x$ and $z$ denote horizontal position and altitude in an inertial reference frame, respectively, and $\theta$ is the pitch angle.
The state is ${\boldsymbol{x} \in \mathcal{X} \subset \mathbb{R}^2 \times \mathbb{S}^1 \times \mathbb{R}^3}$ and the inputs are the thrust
${F \in [0,F_{\rm max}]}$ and moment ${M \in [-M_{\rm max}, M_{\rm max}]}$ applied by the propellers. Here, ${g_{\rm D} \!=\! 9.81 \nspace{2}{\rm m/s^2}}$ is the acceleration due to gravity, ${m \!=\! {\rm 1 \nspace{2} kg}}$ is the mass of the quadrotor,
and ${J \!=\! {\rm 0.25 \nspace{2} kg \nspace{1} m^2}}$ is the principal moment of inertia about the $y$-axis. The components of the disturbance are given by ${d_4(t) \!=\! 1 \nspace{2}{\rm m/s^2}}$ and ${d_5(t) \!=\! \frac{1}{2}{\rm sin}(0.3t \!-\! \frac{\pi}{3})\nspace{2}{\rm m/s^2}}$. 

We consider the motivating case where a human operator loses connection with the quadrotor~\cite{singletary2021onboard}, such that ${\boldsymbol{k}_{\rm p}(\boldsymbol{x}) = \boldsymbol{0}}$, and the controller must prevent crashing into the ground.
The safe set ${\Cs \triangleq \{\boldsymbol{x} \in \mathcal{X} \!:\! h(\x) \!=\! z \!-\! z_{\rm min} \!\geq\! 0 \}}$  is thus defined by a minimum altitude ${z_{\rm min} > 0}$.
The backup controller 
${\boldsymbol{k}_{\rm b}(\x) = [F_{\rm max},\nspace{2} K_{\rm p} \theta \!+\! K_{\rm d} \dot{\theta}]^\top}$,
with gains ${K_{\rm p}, K_{\rm d} > 0}$, aims to bring the quadrotor to horizontal and apply maximum thrust to prevent a crash. The backup set is defined by ${\Cb \triangleq \{\boldsymbol{x} \in \mathcal{X} : h_{\rm b}(\x),  h(\x) \geq 0 \}}$.
The function ${h_{\rm b}(\x) \!=\! -\frac{1}{\kappa} \ln \big( {\rm e}^{-\kappa h_1(\x)} \!+\! {\rm e}^{-\kappa h_2(\x)}\big)}$ with ${\kappa > 0}$ under-approximates ${\min \{ h_1(\x), h_2(\x)\}}$, as in \cite{lindemann_stlcbf_2019,tamas_composing23},
with
${h_1(\x) = \dot{z}}$,
${h_2(\x) = 1 - (K_{\rm p}\theta^2 + J\dot{\theta}^2)/(2c_{\rm b})}$, where ${c_{\rm b} > 0}$. It can be shown that $\boldsymbol{k}_{\rm b}$ renders $\Cb$ robustly forward invariant and satisfies input constraints if ${c_{\rm b} \! \leq \!
\frac{K_p \theta_{\max}^2}{2}}$, ${c_{\rm b} \!\leq\!
\frac{J \dot\theta_{\max}^2}{2}}$, ${- g_D
\!+\! \frac{F_{\max}}{m} \cos\left(\sqrt{\frac{2c_{\rm b}}{K_p}}\right) \!- \!\delta_d
\!\geq\! 0}$, and ${2c_{\rm b} \left(
K_p\! + \!\frac{K_d^2}{J}
\right) \!\leq\! M_{\max}^2}$ for ${\frac{\pi}{2} \!>\! \theta_{\rm max} \!>\! 0, \dot{\theta}_{\rm max} \!>\! 0}$. We omit the proof for brevity\footnote{See \href{https://github.com/davidvwijk/UE-bCBF}{https://github.com/davidvwijk/UE-bCBF} for  additional details.}. For system~\eqref{eq:quadrotor} under $\ub$, the closed-loop Jacobian depends only on constant parameters and $\theta$. Therefore, one can obtain $c$ such that ${\mu(\jac(\tau,\x)) \!\leq\! c}$, allowing the flow bound in Lemma~\ref{lemma: logBound} to be used.

Fig.~\ref{fig:quadrotor} shows the simulation results for system~\eqref{eq:quadrotor} with the proposed~\eqref{eq:dob-bcbf-qp} controller\footnote{The simulation uses
${F_{\rm max} = 20 \nspace{2}{\rm N}}$,
${M_{\rm max} = 20 \nspace{2}{\rm Nm}}$,
${K_{\rm p} = 1 \nspace{2}{\rm Nm}}$,
${K_{\rm d} = 1.01 \nspace{2}{\rm Nms}}$,
${\kappa = 5}$,
${\theta_{\rm max} = 55\deg}$, and
${\dot{\theta}_{\rm max} = 3 \nspace{2}{\rm rad/s}}$.
}. In the simulation, 
similar behavior is observed for the nonlinear and higher-dimensional quadrotor dynamics as for the double integrator.
The proposed controller ensures robust safety, i.e., prevents the quadrotor from crashing, even in the presence of disturbances while satisfying input constraints.
This behavior is achieved using an estimate of the disturbance, which is improved over time via the uncertainty estimator~\eqref{eq:DOB_all}.

\begin{figure}
\centering
{\centerline{\includegraphics[width=1.0\columnwidth]{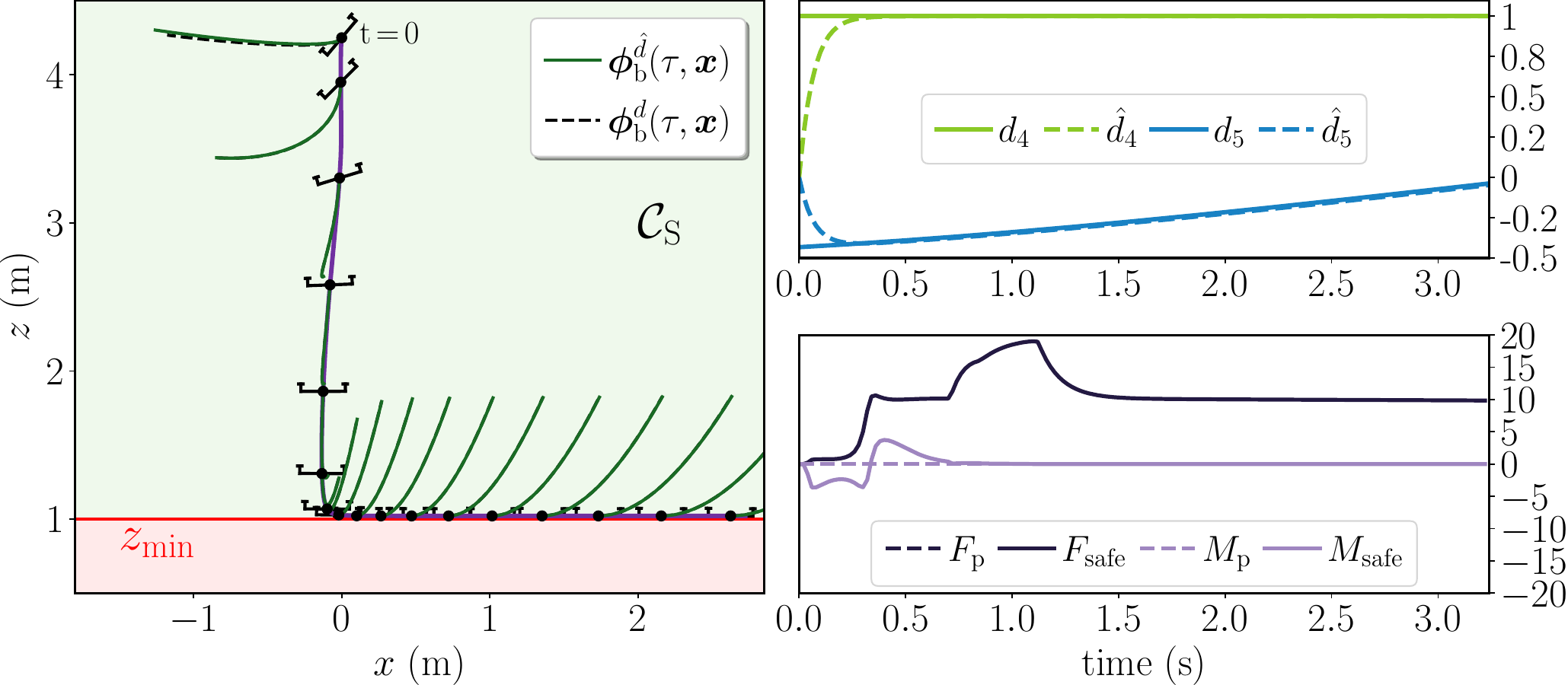}}}
\vspace{.2cm}
\caption{Simulation of the quadrotor~\eqref{eq:quadrotor} using the proposed uncertainty estimator backup CBF controller \eqref{eq:dob-bcbf-qp}. The trajectory of the system (purple) indicates safe behavior despite the unknown disturbance (\textbf{left}). The controller uses the estimated backup trajectories (green) that approach the unknown backup trajectories under the true disturbance (black dashed).
The disturbance estimate converges to the true value (\textbf{top right}), while the control inputs stay within the prescribed bounds (\textbf{bottom right}). See animation at \href{https://youtu.be/btNq8rAtAkM}{\footnotesize\texttt{https://youtu.be/btNq8rAtAkM}}.}
\label{fig:quadrotor}
\end{figure}

\end{casestudy}

%% file: sections/06_Conclusion.tex
\section{Conclusion}\label{sec:conclusion}

We presented a novel framework to guarantee online controlled invariance in the presence of unknown bounded disturbances for input-constrained systems.
We developed a framework that enables the utilization of a broad class of uncertainty estimators to reduce conservatism by reconstructing an estimate of the unknown disturbance. We provided forward invariance conditions for a subset of a controlled invariant set considering the disturbed system, and proved that enforcing these conditions guarantees safety for the disturbed system. Further, our approach accounts for generalized estimate sensitivity terms, thus providing substantially more value than simply merging existing safety-critical control techniques.

%% file: sections/07_Appendix.tex
\section*{Appendix}
\renewcommand{\thesubsection}{\Alph{subsection}}

\subsection{Proof of \Cref{lemma:total_derivs}}
\begin{proof} \label{proof:total_derivs}
        By the chain rule, we have 
        \begin{align*}
            & \dot{h}(\phidhat, \boldsymbol{u})  \\
            &= \frac{\partial h}{\partial \phidhat}  \frac{\partial \phidhat}{\partial \x} {\frac{\partial \x}{\partial t}} + \frac{\partial h}{\partial \phidhat}  \frac{\partial \phidhat}{\partial \dhat} {\frac{\partial \dhat}{\partial t}} \\
            &= \nabla h(\phidhat) \boldsymbol{\Phi}(\tau,\x) \dot{\x} + \nabla h(\phidhat) \boldsymbol{\Theta}(\tau,\x) \dhatdot.
        \end{align*}
        To evaluate the sensitivity terms, we use the knowledge of the estimate flow evolution over the backup horizon: 
        \begin{align*}
            \frac{\partial}{\partial \x} \Big( \frac{\partial }{\partial \tau}\phidhat \Big)
            &= \frac{\partial}{\partial \x} \Big( \boldsymbol{f}_{\rm cl}(\boldsymbol{\phi}^{\hat{d}}_{\rm b}(\tau,\x)) + \dhat(t) \Big) \\
            &= \frac{\partial \boldsymbol{f}_{\rm cl}}{\partial \phidhat}\frac{\partial \phidhat}{\partial \x} \\
            &= \jac(\phidhat) \frac{\partial \phidhat}{\partial \x},
        \end{align*}
        where ${\jac}$ is the closed-loop Jacobian from \eqref{eq:F_cl_def} evaluated at $\phidhat$.
        Because $\x$ and $\tau$ are independent variables, ${\frac{\partial}{\partial \x} \big(\nspace{-2}  \frac{\partial }{\partial \tau}\phidhat \big) \!=\! \frac{\partial}{\partial \tau} \big( \nspace{-2} \frac{\partial }{\partial \x}\phidhat \big)}$ and thus we have
        \begin{align*} 
            \frac{\partial}{{\partial} \tau} {{\boldsymbol{\Phi}}}(\tau, \boldsymbol{x}) = \jac(\phidhat) {{\boldsymbol{\Phi}}}(\tau, \boldsymbol{x}), \sspace
            \boldsymbol{\Phi}(0,\boldsymbol{x}) = {\boldsymbol{I}},
        \end{align*}
        where 
        the initial condition is obtained by differentiating the initial condition ${\boldsymbol{\phi}^{\hat{d}}_{\rm b}(0,\x) = \x}$ w.r.t. $\x$; cf. \eqref{eq:est_dist_flow}. Thus, we forward integrate ${\boldsymbol{\Phi}(\tau,\x)}$ using \eqref{eq:dobcbf_stmProp} along with the flow. 
        Similarly, we can obtain the sensitivity of the estimate flow to changes in the disturbance estimate at time $t$: 
        \begin{align*}
            \frac{\partial}{\partial \dhat} \Big( \frac{\partial }{\partial \tau}\phidhat \Big) 
            &= \frac{\partial}{\partial \dhat} \Big( \boldsymbol{f}_{\rm cl}(\boldsymbol{\phi}^{\hat{d}}_{\rm b}(\tau,\x)) + \dhat(t) \Big) \\
            &= \frac{\partial \boldsymbol{f}_{\rm cl}}{\partial \phidhat}\frac{\partial \phidhat}{\partial \dhat} + {\frac{\partial \dhat(t)}{\partial \dhat}} \\
            &= \jac(\phidhat) \frac{\partial \phidhat}{\partial \dhat} + \eye.
        \end{align*}
        We have ${\frac{\partial}{\partial \dhat}  \nspace{-2} \big( \nspace{-2} \frac{\partial }{\partial \tau}\phidhat \nspace{-2}\big ) \nspace{-2} \!=\!  \nspace{-2} \frac{\partial}{\partial \tau} \nspace{-2} \big( \nspace{-2} \frac{\partial }{\partial \dhat}\phidhat \nspace{-2} \big )}$ since $\tau$ and $\dhat$ are independent, 
        and thus ${\boldsymbol{\Theta}(\tau,\x)}$
        evolves with
        \begin{align*} 
            \frac{\partial}{{\partial} \tau}{{\boldsymbol{\Theta}}}(\tau, \boldsymbol{x}) \! = \! \jac(\phidhat)\boldsymbol{\Theta}(\tau,\x) \!+ \!\eye,
            \nspace{6}
            \boldsymbol{\Theta}(0,\boldsymbol{x}) \! = \! {\boldsymbol{0}}, 
        \end{align*}
        where the initial condition is obtained by differentiating the initial condition ${\boldsymbol{\phi}^{\hat{d}}_{\rm b}(0,\x) = \x}$ w.r.t.~$\dhat$. Performing a similar analysis for ${\dot{h}_{\rm b}(\phidhatT, \boldsymbol{u})}$ completes the proof.
\end{proof}

%% file: bib/references.bib
@article{jankovic_robust_2018,
	title = {Robust control barrier functions for constrained stabilization of nonlinear systems},
	volume = {96},
	issn = {00051098},
	url = {https://linkinghub.elsevier.com/retrieve/pii/S0005109818303509},
	doi = {10.1016/j.automatica.2018.07.004},
	language = {en},
	urldate = {2024-03-18},
	journal = {Automatica},
	author = {Jankovic, Mrdjan},
	year = {2018},
	pages = {359--367},
}

@inproceedings{garg_robust_2021,
	title = {Robust {Control} {Barrier} and {Control} {Lyapunov} {Functions} with {Fixed}-{Time} {Convergence} {Guarantees}},
	isbn = {978-1-66544-197-1},
	url = {https://ieeexplore.ieee.org/document/9482751/},
	doi = {10.23919/ACC50511.2021.9482751},
	language = {en},
	urldate = {2024-03-18},
	booktitle = {Proc. Amer. Control Conf.},
	author = {Garg, Kunal and Panagou, Dimitra},
	year = {2021},
	pages = {2292--2297},
}

@article{shaw_cortez_control_2021,
	title = {Control {Barrier} {Functions} for {Mechanical} {Systems}: {Theory} and {Application} to {Robotic} {Grasping}},
	volume = {29},
	issn = {1063-6536, 1558-0865, 2374-0159},
	shorttitle = {Control {Barrier} {Functions} for {Mechanical} {Systems}},
	url = {https://ieeexplore.ieee.org/document/8913716/},
	doi = {10.1109/TCST.2019.2952317},
	language = {en},
	number = {2},
	urldate = {2024-03-18},
	journal = {IEEE Trans. Control Syst. Technol.},
	author = {Shaw Cortez, Wenceslao and Oetomo, Denny and Manzie, Chris and Choong, Peter},
	year = {2021},
	pages = {530--545},
}

@INPROCEEDINGS{issf_og,
  author={Romdlony, Muhammad Zakiyullah and Jayawardhana, Bayu},
  booktitle={55th IEEE Conf. Decision and Control}, 
  title={On the new notion of input-to-state safety}, 
  year={2016},
  volume={},
  number={},
  pages={6403-6409},
  doi={10.1109/CDC.2016.7799254}}

@ARTICLE{Issf_ames,
  author={Kolathaya, Shishir and Ames, Aaron D.},
  journal={IEEE Control Syst. Lett.}, 
  title={Input-to-State Safety With Control Barrier Functions}, 
  year={2019},
  volume={3},
  number={1},
  pages={108-113},
  doi={10.1109/LCSYS.2018.2853698}}

@ARTICLE{lopez_adaptive_robustCBF,
  author={Lopez, Brett T. and Slotine, Jean-Jacques E. and How, Jonathan P.},
  journal={IEEE Control Syst. Lett.}, 
  title={Robust Adaptive Control Barrier Functions: An Adaptive and Data-Driven Approach to Safety}, 
  year={2021},
  volume={5},
  number={3},
  pages={1031-1036},
  doi={10.1109/LCSYS.2020.3005923}}

@ARTICLE{lindemann2024learningrobustoutputcontrol,
  author={Lindemann, Lars and Robey, Alexander and Jiang, Lejun and Das, Satyajeet and Tu, Stephen and Matni, Nikolai},
  journal={IEEE Open J. Control Syst.}, 
  title={Learning Robust Output Control Barrier Functions From Safe Expert Demonstrations}, 
  year={2024},
  volume={3},
  number={},
  pages={158-172},
  doi={10.1109/OJCSYS.2024.3385348}}

@ARTICLE{emam_data_driven,
  author={Emam, Yousef and Glotfelter, Paul and Wilson, Sean and Notomista, Gennaro and Egerstedt, Magnus},
  journal={IEEE Trans. on Robotics}, 
  title={Data-Driven Robust Barrier Functions for Safe, Long-Term Operation}, 
  year={2022},
  volume={38},
  number={3},
  pages={1671-1685},
  doi={10.1109/TRO.2021.3118965}}

@inproceedings{gurriet_online_2018,
	title = {An {Online} {Approach} to {Active} {Set} {Invariance}},
	isbn = {978-1-5386-1395-5},
	url = {https://ieeexplore.ieee.org/document/8619139/},
	doi = {10.1109/CDC.2018.8619139},
	language = {en},
	urldate = {2023-11-28},
	booktitle = {Proc. 57th IEEE Conf. Decision and Control},
	author = {Gurriet, Thomas and Mote, Mark and Ames, Aaron D. and Feron, Eric},
	year = {2018},
	pages = {3592--3599},
}

@ARTICLE{gurriet_scalable_2020,
  author={Gurriet, Thomas and Mote, Mark and Singletary, Andrew and Nilsson, Petter and Feron, Eric and Ames, Aaron D.},
  journal={IEEE Access}, 
  title={A Scalable Safety Critical Control Framework for Nonlinear Systems}, 
  year={2020},
  volume={8},
  number={},
  pages={187249-187275},
  doi={10.1109/ACCESS.2020.3025248}}

@Article{singletary2021onboard,
  author={Andrew Singletary and Aiden Swann and Yuxiao Chen and Aaron D. Ames},
  journal = {IEEE Robot. Autom. Lett.},
  title={Onboard Safety Guarantees for Racing Drones: High-speed Geofencing with Control Barrier Functions},
  year={2022},
  volume={7},
  number={2},
  pages={2897--2904},
}

@book{khalil2002nonlinear,
  title={Nonlinear Systems},
  author={Khalil, H.K.},
  isbn={9780130673893},
  lccn={95045804},
  series={Pearson Education},
  url={https://books.google.com/books?id=t_d1QgAACAAJ},
  year={2002},
  edition={2},
  publisher={Prentice Hall}
}

@ARTICLE{ames_2017,
  author={Ames, Aaron D. and Xu, Xiangru and Grizzle, Jessy W. and Tabuada, Paulo},
  journal={IEEE Trans. Autom. Control}, 
  title={Control Barrier Function Based Quadratic Programs for Safety Critical Systems}, 
  year={2017},
  volume={62},
  number={8},
  pages={3861-3876},
  doi={10.1109/TAC.2016.2638961}}

@Book{contraction_bullo,
  author =    {F. Bullo},
  title =     {Contraction Theory for Dynamical Systems},
  year =      2023,
  edition =   {{1.1}},
  publisher = {Kindle Direct Publishing},
  ISBN =      {979-8836646806},
  url =       {https://fbullo.github.io/ctds},
}

@Inbook{Sontag2010,
author="Sontag, Eduardo D.",
title="Contractive Systems with Inputs",
bookTitle="Perspectives in Mathematical System Theory, Control, and Signal Processing: A Festschrift in Honor of Yutaka Yamamoto on the Occasion of his 60th Birthday",
year="2010",
publisher="Springer Berlin Heidelberg",
address="Berlin, Heidelberg",
pages="217--228",
isbn="978-3-540-93918-4",
doi="10.1007/978-3-540-93918-4_20",
url="https://doi.org/10.1007/978-3-540-93918-4_20"
}

@article{LOHMILLER1998683,
title = {On Contraction Analysis for Non-linear Systems},
journal = {Automatica},
volume = {34},
number = {6},
pages = {683-696},
year = {1998},
issn = {0005-1098},
doi = {https://doi.org/10.1016/S0005-1098(98)00019-3},
url = {https://www.sciencedirect.com/science/article/pii/S0005109898000193},
author = {Winfried Lohmiller and Jean-Jacques E. Slotine}
}

@misc{vanwijk2026outputfeedbackbackupcontrol,
      title={Output Feedback Backup Control Barrier Functions: Safety Guarantees Under Input Bounds and State Estimation Error}, 
      author={David E. J. van Wijk and Tamas G. Molnar and Samuel Coogan and Manoranjan Majji and Aaron D. Ames and Joel W. Burdick},
      year={2026},
      eprint={2604.19893},
      archivePrefix={arXiv},
      primaryClass={eess.SY},
      url={https://arxiv.org/abs/2604.19893}, 
}

@ARTICLE{9640564,
  author={Xie, Huahui and Dai, Li and Lu, Yuchen and Xia, Yuanqing},
  journal={IEEE Trans. Autom. Control}, 
  title={Disturbance Rejection {MPC} Framework for Input-Affine Nonlinear Systems}, 
  year={2022},
  volume={67},
  number={12},
  pages={6595-6610},
  doi={10.1109/TAC.2021.3133376}}

@article{DU2016207,
title = {Robust dynamic positioning of ships with disturbances under input saturation},
journal = {Automatica},
volume = {73},
pages = {207-214},
year = {2016},
issn = {0005-1098},
doi = {https://doi.org/10.1016/j.automatica.2016.06.020},
url = {https://www.sciencedirect.com/science/article/pii/S0005109816302473},
author = {Jialu Du and Xin Hu and Miroslav Krsti\'c and Yuqing Sun}
}

@ARTICLE{contraction_lcss_bullo24,
  author={Davydov, Alexander and Bullo, Francesco},
  journal={IEEE Control Syst. Lett.}, 
  title={Perspectives on Contractivity in Control, Optimization, and Learning}, 
  year={2024},
  volume={8},
  number={},
  pages={2087-2098},
  doi={10.1109/LCSYS.2024.3436127}}

@book{blanchini_set-theoretic_2015,
	address = {Cham},
	series = {Systems \& {Control}: {Foundations} \& {Applications}},
	title = {Set-{Theoretic} {Methods} in {Control}},
	copyright = {https://www.springernature.com/gp/researchers/text-and-data-mining},
    edition = {Second},
	isbn = {978-3-319-17932-2 978-3-319-17933-9},
	url = {https://link.springer.com/10.1007/978-3-319-17933-9},
	language = {en},
	urldate = {2024-11-14},
	publisher = {Springer International Publishing},
	author = {Blanchini, Franco and Miani, Stefano},
	year = {2015},
	doi = {10.1007/978-3-319-17933-9},
}

@ARTICLE{lindemann_stlcbf_2019,
  author={Lindemann, Lars and Dimarogonas, Dimos V.},
  journal={IEEE Control Syst. Lett.}, 
  title={Control Barrier Functions for Signal Temporal Logic Tasks}, 
  year={2019},
  volume={3},
  number={1},
  pages={96-101},
  doi={10.1109/LCSYS.2018.2853182}}

@ARTICLE{tamas_composing23,
  author={Molnar, Tamas G. and Ames, Aaron D.},
  journal={IEEE Control Syst. Lett.}, 
  title={Composing Control Barrier Functions for Complex Safety Specifications}, 
  year={2023},
  volume={7},
  number={},
  pages={3615-3620},
  doi={10.1109/LCSYS.2023.3339719}}

@INPROCEEDINGS{wu_acc_quadrotor2016,
  author={Wu, Guofan and Sreenath, Koushil},
  booktitle={Proc. Amer. Control Conf.}, 
  title={Safety-critical control of a planar quadrotor}, 
  year={2016},
  volume={},
  number={},
  pages={2252--2258},
  doi={10.1109/ACC.2016.7525253}}

@ARTICLE{vanWijk_DRbCBF_24,
  author={van Wijk, David E. J. and Coogan, Samuel and Molnar, Tamas G. and Majji, Manoranjan and Hobbs, Kerianne L.},
  journal={IEEE Control Syst. Lett.}, 
  title={Disturbance-Robust Backup Control Barrier Functions: Safety Under Uncertain Dynamics}, 
  year={2024},
  volume={8},
  number={},
  pages={2817-2822},
  doi={10.1109/LCSYS.2024.3514998}
}

@ARTICLE{UE_survey_2016,
  author={Chen, Wen-Hua and Yang, Jun and Guo, Lei and Li, Shihua},
  journal={IEEE Trans. Ind. Electron.}, 
  title={Disturbance-Observer-Based Control and Related Methods{---}{A}n Overview}, 
  year={2016},
  volume={63},
  number={2},
  pages={1083-1095},
  doi={10.1109/TIE.2015.2478397}}

@article{rabiee_softmin_bcbf,
title = {Soft-minimum and soft-maximum barrier functions for safety with actuation constraints},
journal = {Automatica},
volume = {171},
pages = {111921},
year = {2025},
issn = {0005-1098},
doi = {https://doi.org/10.1016/j.automatica.2024.111921},
url = {https://www.sciencedirect.com/science/article/pii/S0005109824004151},
author = {Pedram Rabiee and Jesse B. Hoagg}
}

@article{vanwijk2025safetycriticalcontrolboundedinputs,
  title={Safety-Critical Control with Bounded Inputs: A Closed-Form Solution for Backup Control Barrier Functions},
  author={van Wijk, David E. J. and Da{\c{s}}, Ersin and Molnar, Tamas G and Ames, Aaron D and Burdick, Joel W},
  journal={arXiv preprint arXiv:2510.05436},
  year={2025}
}

@ARTICLE{das2025robustcontrolbarrierfunctions,
  author={Da{\c{s}}, Ersin and Burdick, Joel W.},
  journal={IEEE Trans. Autom. Control}, 
  title={Robust Control Barrier Functions Using Uncertainty Estimation With Application to Mobile Robots}, 
  year={2025},
  volume={70},
  number={7},
  pages={4766-4773},
  doi={10.1109/TAC.2025.3538742}}

@INPROCEEDINGS{wang2023DOB,
  author={Wang, Yujie and Xu, Xiangru},
  booktitle={Proc. Amer. Control Conf.}, 
  title={Disturbance Observer-based Robust Control Barrier Functions}, 
  year={2023},
  volume={},
  number={},
  pages={3681-3687},
  doi={10.23919/ACC55779.2023.10156095}}

@ARTICLE{Isaly2024DOB,
  author={Isaly, Axton and Patil, Omkar Sudhir and Sweatland, Hannah M. and Sanfelice, Ricardo G. and Dixon, Warren E.},
  journal={IEEE Trans. Autom. Control}, 
  title={Adaptive Safety with a RISE-Based Disturbance Observer}, 
  year={2024},
  volume={69},
  number={7},
  pages={4883-4890},
  doi={10.1109/TAC.2024.3358210}}

@ARTICLE{alan2022tunableissf,
  author={Alan, Anil and Taylor, Andrew J. and He, Chaozhe R. and Orosz, Gábor and Ames, Aaron D.},
  journal={IEEE Control Syst. Lett.}, 
  title={Safe Controller Synthesis With Tunable Input-to-State Safe Control Barrier Functions}, 
  year={2022},
  volume={6},
  number={},
  pages={908-913},
  doi={10.1109/LCSYS.2021.3087443}}

@article{yan2023surviving,
  title={Surviving disturbances: A predictive control framework with guaranteed safety},
  author={Yan, Yunda and Wang, Xue-Fang and Marshall, Benjamin James and Liu, Cunjia and Yang, Jun and Chen, Wen-Hua},
  journal={Automatica},
  volume={158},
  pages={111238},
  year={2023},
  publisher={Elsevier}
}
